\documentclass[draft]{agujournal2019}%[draft]
\usepackage{url} %this package should fix any errors with URLs in refs.
\usepackage{lineno}
\usepackage{booktabs}
\usepackage{comment}
\usepackage{amsmath,amsthm,amssymb}
\usepackage[normalem]{ulem}
\usepackage[inline]{trackchanges} %for better track changes. finalnew option will compile document with changes incorporated.

\draftfalse

\journalname{Space Weather}

\begin{document}

%% ------------------------------------------------------------------------ %%
%  Title
%
% (A title should be specific, informative, and brief. Use
% abbreviations only if they are defined in the abstract. Titles that
% start with general keywords then specific terms are optimized in
% searches)
%
%% ------------------------------------------------------------------------ %%

% Example: \title{This is a test title}

\title{Solar wind prediction using deep learning}

%% ------------------------------------------------------------------------ %%
%
%  AUTHORS AND AFFILIATIONS
%
%% ------------------------------------------------------------------------ %%

% Authors are individuals who have significantly contributed to the
% research and preparation of the article. Group authors are allowed, if
% each author in the group is separately identified in an appendix.)

% List authors by first name or initial followed by last name and
% separated by commas. Use \affil{} to number affiliations, and
% \thanks{} for author notes.
% Additional author notes should be indicated with \thanks{} (for
% example, for current addresses).

% Example: \authors{A. B. Author\affil{1}\thanks{Current address, Antartica}, B. C. Author\affil{2,3}, and D. E.
% Author\affil{3,4}\thanks{Also funded by Monsanto.}}

\authors{Vishal Upendran\affil{1,2}, Mark C. M. Cheung\affil{3,4}, Shravan Hanasoge\affil{5}, \\Ganapathy Krishnamurthi\affil{1}}

 \affiliation{1}{Inter-University Centre for Astronomy and Astrophysics, Pune, India}
 \affiliation{2}{Department of Engineering Design, Indian Institute of Technology- Madras, Chennai, India}
 \affiliation{3}{Lockheed Martin Solar and Astrophysics Laboratory, Palo Alto, CA 94304, USA}
 \affiliation{4}{Hansen Experimental Physics Laboratory, Stanford University, Stanford, CA 94305, USA}
 \affiliation{5}{Department of Astronomy and Astrophysics, Tata Institute of Fundamental Research, Mumbai, India}

%(repeat as many times as is necessary)

%% Corresponding Author:
% Corresponding author mailing address and e-mail address:

% (include name and email addresses of the corresponding author.  More
% than one corresponding author is allowed in this LaTeX file and for
% publication; but only one corresponding author is allowed in our
% editorial system.)

% Example: \correspondingauthor{First and Last Name}{email@address.edu}

%\correspondingauthor{=name=}{=email address=}

%% Keypoints, final entry on title page.

%  List up to three key points (at least one is required)
%  Key Points summarize the main points and conclusions of the article
%  Each must be 100 characters or less with no special characters or punctuation and must be complete sentences

% Example:
% \begin{keypoints}
% \item	List up to three key points (at least one is required)
% \item	Key Points summarize the main points and conclusions of the article
% \item	Each must be 100 characters or less with no special characters or punctuation and must be complete sentences
% \end{keypoints}

\begin{keypoints}
\item Machine learning models are evaluated for the prediction of solar wind (SW) speed measured at Lagrangian point 1 between the Sun and Earth.
\item Without imposing physics rules, deep neural networks (DNNs) can be trained on Extreme UV images of the solar corona to forecast SW speed.
\item Our DNN WindNet significantly outperforms simpler models, and pays attention to coronal holes in 193 {\AA} for fast wind prediction.
%\item The DNN considered here, named WindNet, significantly outperforms simpler models (e.g. 27 day persistence), especially when the look-forward time exceeds one day.
%\item Visualization of node activations in WindNet reveals that the DNN pays attention to coronal holes (CHs) in 193 {\AA} EUV images when forecasting a fast solar wind. %Conversely, the DNN pays attention to non-CH regions when forecasting slow solar wind speed.
\end{keypoints}

%% ------------------------------------------------------------------------ %%
%
%  ABSTRACT and PLAIN LANGUAGE SUMMARY
%
% A good Abstract will begin with a short description of the problem
% being addressed, briefly describe the new data or analyses, then
% briefly states the main conclusion(s) and how they are supported and
% uncertainties.

% The Plain Language Summary should be written for a broad audience,
% including journalists and the science-interested public, that will not have 
% a background in your field.
%
% A Plain Language Summary is required in GRL, JGR: Planets, JGR: Biogeosciences,
% JGR: Oceans, G-Cubed, Reviews of Geophysics, and JAMES.
% see http://sharingscience.agu.org/creating-plain-language-summary/)
%
%% ------------------------------------------------------------------------ %%

%% \begin{abstract} starts the second page

\begin{abstract}
Emanating from the base of the Sun's corona, the solar wind fills the interplanetary medium with a magnetized stream of charged particles whose interaction with the Earth's magnetosphere has space-weather consequences such as geomagnetic storms. Accurately predicting the solar wind through measurements of the spatio-temporally evolving conditions in the solar atmosphere is important but remains an unsolved problem in heliophysics and space-weather research.

In this work, we use deep learning for prediction of solar wind (SW) properties. We use Extreme Ultraviolet images of the solar corona from space based observations %the Atmospheric Imaging Assembly on board NASA's Solar Dynamics Observatory 
to predict the SW speed from the NASA OMNIWEB dataset, measured at Lagragian point 1. We evaluate our model against autoregressive and naive models, and find that our model outperforms the benchmark models, obtaining a best-fit correlation of 0.55 $\pm$ 0.03 with the observed data.

%Upon visualization and investigation of how the model uses data to make predictions, we find that our proposed model has higher activation at the coronal holes closer to the day of prediction, and at the active regions farther away from the day of prediction. This trend of activation bears an uncanny similarity to the influence of regions potentially being the sources of fast and slow wind, as reported in existing literature.
Upon visualization and investigation of how the model uses data to make predictions, we find higher activation at the coronal holes for fast wind prediction ($\approx$ 3 to 4 days prior to prediction), and at the active regions for slow wind prediction. These trends bear an uncanny similarity to the influence of regions potentially being the sources of fast and slow wind, as reported in literature.
This suggests that our model was able to learn some of the salient associations between coronal and solar wind structure without built-in physics knowledge. Such an approach may help us discover hitherto unknown relationships in heliophysics datasets.
\end{abstract}

\section*{Plain Language Summary}
%[ enter your Plain Language Summary here or delete this section]
The solar wind is a stream of particles coming from the Sun.
The interaction of the solar wind with the Earth's magnetosphere gives rise to space weather effects, including geomagnetic storms, aurorae and disruptions to electrical distribution grids. Accurate prediction of the solar wind is of interest to government agencies and private industry.
%Such particles when they interact with the Earth's magnetic field, give rise to lasting space-weather effects like aurora and geomagnetic storms. Thus, accurately predicting the properties of the solar wind becomes very important.
In this work, we explore the use of machine learning models to predict the solar wind speed as measured at the Lagrangian point 1 (L1) between the Sun and Earth. The best performing method is a deep neural network that uses Extreme Ultraviolet (EUV) imagery data from NASA's Solar Dynamics Observatory (SDO) as input. Without explicitly building in physical relationships into the model, it is able to outperform a number of baseline models. We find the model pays attention to regions on the Sun that are in agreement with heuristics used in the literature (e.g. coronal holes for the fast solar wind). Such an approach may, in the future, help us discover new relationships in heliophysics.

%% ------------------------------------------------------------------------ %%
%
%  TEXT
%
%% ------------------------------------------------------------------------ %%

%%% Suggested section heads:
%
% The main text should start with an introduction. Except for short manuscripts (such as comments and replies), the text should be divided into sections, each with its own heading.
% Headings should be sentence fragments and do not begin with a lowercase letter or number. Examples of good headings are:

% \section{Materials and Methods}
% Here is text on Materials and Methods.
% \subsection{A descriptive heading about methods}
% More about Methods.
% \section{Data} (Or section title might be a descriptive heading about data)
% \section{Results} (Or section title might be a descriptive heading about the
% results)
% \section{Conclusions}

\section{Introduction} \label{sec:intro}
Space weather is defined by the U.S. National Space Weather Plan as the conditions on the sun, in the solar wind (SW), and within Earth's magnetosphere, ionosphere and thermosphere that can influence the performance and reliability of space-borne and ground-based technological systems and can endanger human life or health~\cite{nasa_spaceweather}. The solar wind that influences space weather comprises a continuous stream of magnetized charged particles emanating from the base of the solar corona and permeating the solar system ~\cite{Schwenn2006}. The influence of the SW on space weather arises due to its interaction with the Earth's magnetospehere, resulting in geomagnetic storms and aurorae. Such terrestrial effects of the SW make it extremely important to understand and identify its possible sources, and perform a prediction as a forewarning against potential damages.

\citeA{OwensSWRef} review SW prediction using empirical, physics-based, and hybrid approaches. Typically, a physics-based model uses synoptic magnetograms as the bottom boundary condition. An individual synoptic magnetogram is assembled by sampling the photospheric magnetic flux distribution near the central meridian over the course of a solar rotation. Such magnetograms can be used to extrapolate the surface field into the corona using potential-field source-surface (PFSS)~\cite{AltschulerNewkirk:1969} models or magnetohydrodynamics (MHD) models~\cite<see>[for a comparison between the two]{Riley:MHDvsPFSS}. The global coronal magnetic field (or certain derived properties thereof) may then be used as input for physics-based SW propagation models~\cite<e.g.>[]{SWMHD}, or in the case of a hybrid approach, used for estimation of the SW at L1 using empirical relations. 
WSA-ENLIL and MAS-ENLIL ~\cite{OwensSWRef,Schwenn2006} are among the most widely used SW models. The models provide SW properties such as the velocity, plasma density, magnetic field and temperature. \citeA{jian2015validation} perform a comparison of the various SW models and present a best correlation of 0.57 on hourly prediction using GONG-MAS Thermo-ENLIL model, and 0.50 on the same dataset using GONG-WSA-ENLIL model.

The dependence of high-speed streams on the presence of Coronal Holes (CHs) was first shown in ~\citeA{Krieger1973}, using extrapolation methods.~\citeA{Krieger1973} interpreted this empirical relation as a manifestation of high-speed SW flowing along open magnetic flux regions.~\citeA{wang1990solar} used the inverse of flux tube expansion factor for SW prediction, by identifying an inverse correlation between the flux tube expansion factor and the SW. Recently, there has been a surge in performing regression using fractional (CH) area  extracted from EUV imagery data ~\cite{CorHole_Rotter,MainRefRotterSW,MultiViewpoint}, and a correlations from $\approx$ 0.60 to $\approx$ 0.78 have been obtained between the CH areas and hourly solar wind speed. More recently, ~\citeA{YangNN} devised a Neural network based prediction scheme, taking PFSS model output among other parameters as input, and obtained a correlation of 0.74 on hourly solar wind speed data.

In machine learning (ML) and statistical-learning parlance, the aforementioned traditional empirical models use so-called hand-engineered features as input for their models (e.g. CH area or CH expansion factor). These hand-engineered features are often inspired by some insights from physics-based models, or simply from correlations reported in the literature. Deep learning (DL) is an umbrella term for a broad class of techniques that use neural networks with multiple hidden layers for performing supervised or unsupervised learning tasks. Due to the increased availability of data, and perhaps more importantly, of inexpensive computation, DL has been widely applied in many domains. In areas of engineering and science dealing with ambiguous features, DL has been found to outperform ML algorithms that use hand-engineered features~\cite{Goodfellow-et-al-2016}. Generally, DL involves solving supervised learning tasks such as classification (associating a particular label with a given set of inputs, such as in image classification) or regression (continuous-value prediction). Often, these algorithms need no prior information regarding the exact input-output mapping, but instead try to discover underlying relations by iteratively updating the model parameters by minimizing a loss function (e.g. mean-squared error). Prior information can be built into the model in a number of ways, e.g. (1) by providing hand-engineered input features (which are generally physics-based), (2) assembling a neural net with some layers that have been pre-trained and whose weights are kept fixed during training (known as transfer learning, although the amount of prior information shared is limited by the kind of pretraining performed), and (3) specially customized initialization of weights (usually to accelerate convergence).

Two of the most prominent architectures used in deep learning are Convolutional Neural Networks (ConvNets) and Recurrent Neural Networks (RNNs). ConvNets are a set of deep neural nets which have been successfully applied to different kinds of classification and regression problems~\cite{ciresan2011flexible,deepldefn,lecun2015deep} for image data. These networks work by detecting local patterns at multiple scales in the input, and map them to the appropriate class (classification) or continuous output (regression). RNNs, on the other hand, are a class of deep neural nets designed for understanding the structure of data with a sequential ordering. These have been used extensively for text prediction, natural language processing and regression~\cite{lstm,sutskever2014sequence}.

In this work, we use a ConvNet~\cite{43022} pre-trained on the ImageNet database~\cite{deng2009imagenet}, and couple it with a trainable Long-Short Term Memory cell (LSTM) implementation of an RNN~\cite{lstm}. We use EUV images in 193{\AA} and 211{\AA} from Atmospheric Imaging Assembly (AIA) onboard SDO as input, making features like CHs and ARs clearly visible, and predict the SW speed present in the NASA OMNIWEB dataset. The network is not given any prior information about  the physical mapping between the EUV image data and SW speed, and a direct regression is performed from a time series of AIA images to the SW speed. 
%The network gets a best-fit correlation of $0.55$ between the observed and predicted SW, for 211{\AA} data and $0.51$ for 193{\AA} data. We also find the network shows increased activation at CHs 3 to 4 days prior to prediction, a result consistent with observations and existing literature~\cite{Intro2Plasma,Schwenn_Summ}.
This work is presented as follows: In Sec.~\ref{sec:data} we describe how the input and output data are preprocessed, partitioned into training and test sets, and define some control parameters and evaluation metrics. Then, in Sec.~\ref{sec:MandM}, we briefly introduce the various algorithms used as our benchmarks. We detail our proposed model WindNet, and the visualization technique used for generating activation map. The segmentation algorithms used for the generation of binary masks for the computation of mean activation values, are also described. In Sec.~\ref{sec:result}, we summarize our model predictions vis-a-vis our benchmarks, present the trends of mean activation, and draw conclusions in Sec.~\ref{sec:discuss}. %An introduction to the various ML algorithms used in this work is detailed in the supporting information section for the interested reader.%~\ref{sec:app1} 
%------------------------->
\section{Data and Metrics} \label{sec:data}
\subsection{EUV dataset}
\begin{figure}
\centering 
\leavevmode 
\includegraphics[width=\linewidth]{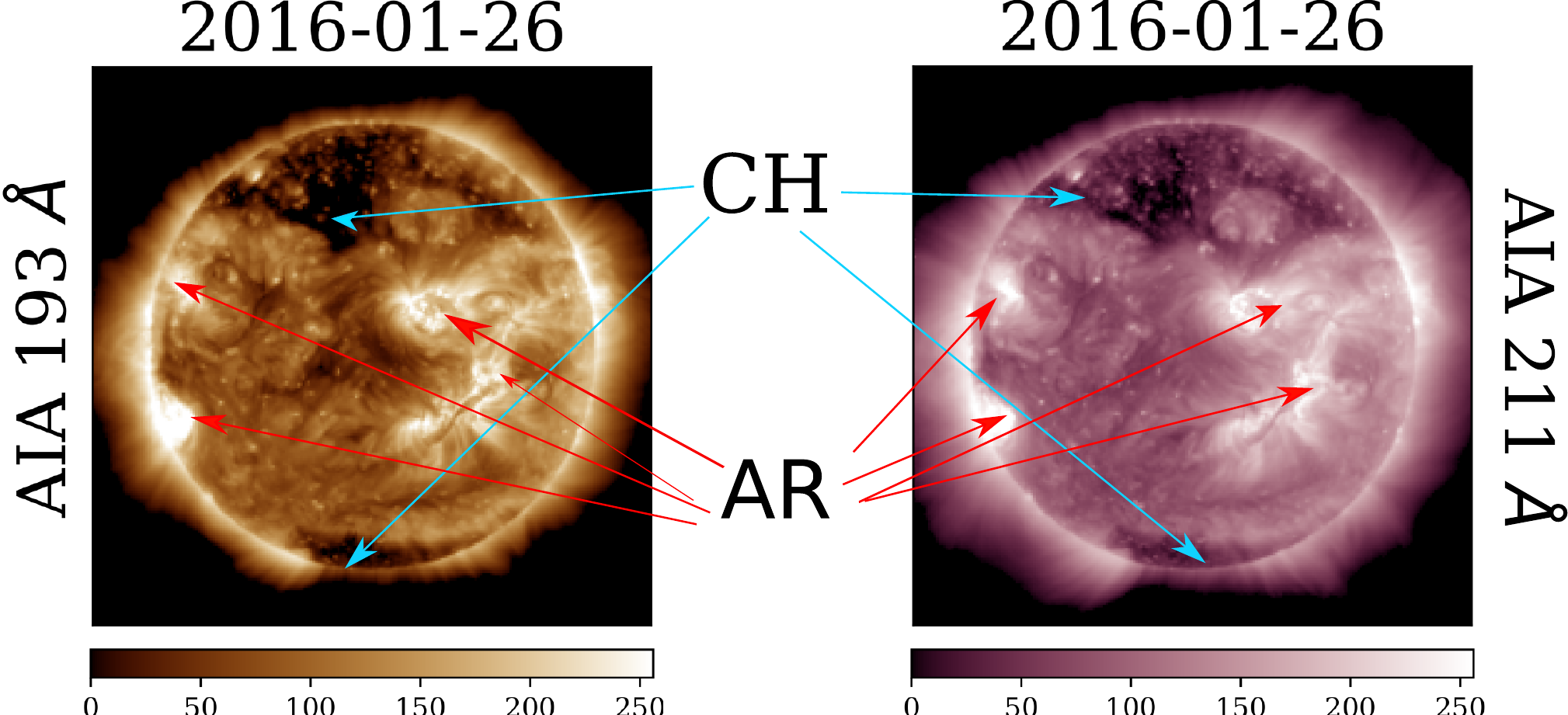}
\caption{Representative AIA data in the 193~{\AA} and 211~{\AA} channels with CHs and ARs marked. This is the final data used in our analysis.}
\label{fig:sdo20}
\end{figure}% observes the Sun
The AIA~\cite{Lemen2012} onboard the SDO~\cite{SDO} is a four-telescope array observing the full Sun in visible, UV and EUV wavelengths. An ML dataset curated from SDO instruments (hereafter, SDOML) has been made publicly available \cite{galvez2019machine}(\url{https://purl.stanford.edu/jc488jb7715} and links therein). AIA images in the dataset have been resampled onto a grid of 512x512 pixels with 4.8 arcsec pixel spacing and are available at 6 min cadence. SDOML images are stored as binary arrays in the Python numpy format~\cite{numpy}. Our training and testing data include AIA images taken at $00:00$ UTC for each day. The selected image forms a proxy for the whole day of observation. However, if the image at $00:00$ does not exist (as is the case with many days), the closest image to $00:00$, from that day, is taken as a proxy for that day.\\
Even during non-flaring times, solar EUV images can have a dynamic range that greatly exceeds the 8-bits per channel dynamic range typical of most computer-vision datasets. For this reason, the input AIA images are first preprocessed by performing log-scaling to bring out fainter features. The images are then passed through a threshold and saturation, which limits the dynamic range of pixel values. This was done to limit the prediction to contribution from Solar disc alone. Furthermore, it was seen that the model performance was better with thresholded and saturated images -- thus, the dynamic range was limited. A general sweep of threshold and saturation was performed for a particular combination of History and delay(\ref{subsec:ControlParam}), for the 193~{\AA} data. The correlation of predicted SW speed with observed SW speed $0.48\pm0.03$ for the best set, with higher thresholds ($\log(250)$,$\log(10000)$) giving us $0.46\pm0.02$ and lower thresholds ($\log(100)$,$\log(1000)$) giving us $0.35\pm0.02$. A coarse search was performed to find the threshold values. The thresholds for 193~{\AA} data were scaled to 211~{\AA} through a ratio of maximum intensities on a given day selected randomly. Equations ~(\ref{eqn:193}) and ~(\ref{eqn:211}) specify the threshold and saturation operations for log scaled 193~{\AA} and 211 {\AA} channel images, respectively. AIA 193~{\AA} and 211~{\AA} data, with CHs and ARs marked, after preprocessing are shown in Fig.~\ref{fig:sdo20}.
%\begin{equation}
%x(94)=
%\begin{cases}
%\log(1.125) & \text{if $x\leq\log(1.25)$} \\
%\log(45.0) & \text{if $x\geq\log(45.0)$} \\
%x & \text{else}
%\end{cases}
%\label{eqn:94}
%\end{equation}
\begin{equation}
x(193)=
\begin{cases}
\log(125.0) & \text{if $x\leq\log(125.0)$} \\
\log(5000.0) & \text{if $x\geq\log(5000.0)$} \\
x & \text{else}
\end{cases}
\label{eqn:193}
\end{equation}
\begin{equation}
x(211)=
\begin{cases}
\log(25.0) & \text{if $x\leq\log(25.0)$} \\
\log(2500.0) & \text{if $x\geq\log(2500.0)$} \\
x & \text{else}
\end{cases}
\label{eqn:211}
\end{equation}
The pixel values are then rescaled between $0.0$ and $255.0$.This is done since our feature extractor expects inputs within this range of values.

%as is the standard range of intensities. This is also done since our feature extractor expects the raw images to have the same range of pixel values.
%-------------------------
\subsection{SW dataset}
\begin{figure}
\centering
\includegraphics[width=\linewidth]{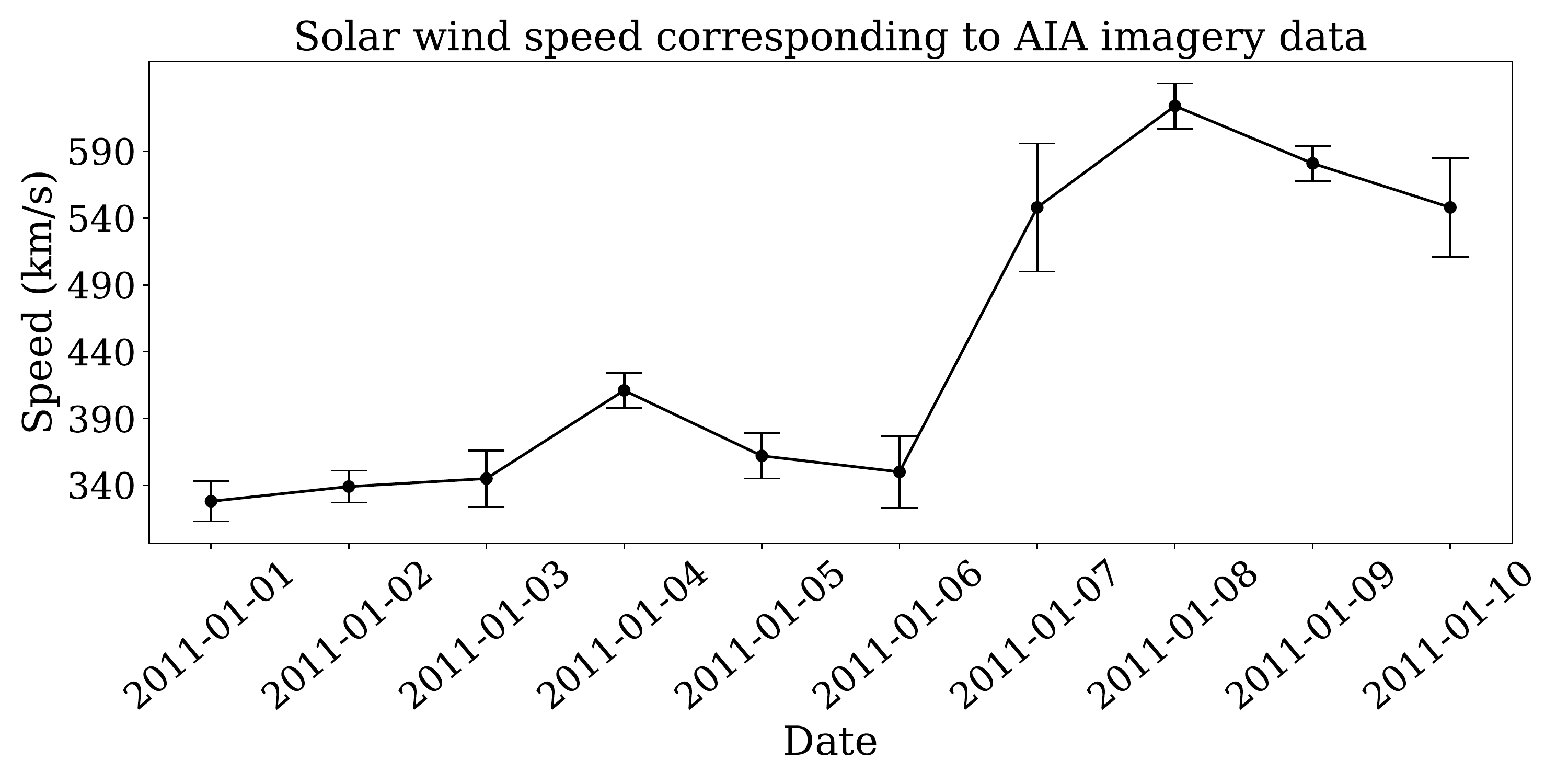}
\caption{10 days of SW speed from NASA OMNIWEB dataset.}
\label{fig:sw1}
\end{figure}

\begin{figure}
    \includegraphics[width=\linewidth]{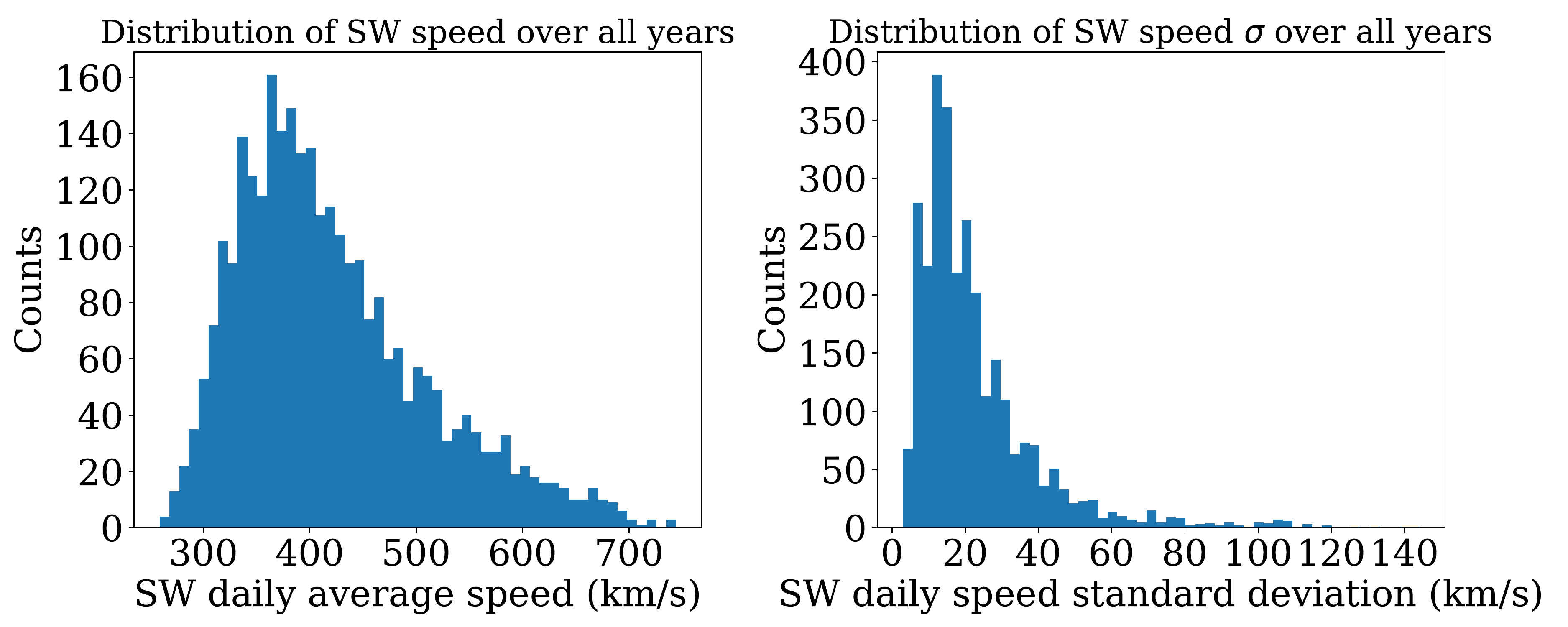}
    \caption{Left: Distribution of SW speed. Right: Distribution of the associated $\sigma$. The distrbutions are computed over the entire dataset.}
    \label{fig:swstats}
\end{figure}

The target output of the prediction models is daily-averaged SW speed measured at L1. Daily averages are used here since the variation of wind speed over a day is not large, and the variation across the mean value sets the uncertainty in wind speed value. The variation (or the standard deviation $\sigma$) is calculated as the variance in hourly measurements over the day, at the OMNIWEB archive (available online at: \url{https://omniweb.gsfc.nasa.gov/form/dx1.html}). %Measurements were retrieved from NASA OMNIWEB archive. %. 
A representative variation in SW speed data over 10 days is plotted in Fig.~\ref{fig:sw1}. The distribution of SW speed and the corresponding $\sigma$ for the entire dataset is shown in Fig.~\ref{fig:swstats}.

%All of our benchmark models are autoregressive -- i.e., they take in a batch of SW speed and give out a corresponding regression value.
There are gaps in the AIA EUV data (30 days of missing data in 211~{\AA} and 31 days of missing data in 193~{\AA}) for \textbf{00:00 UTC}, owing to various reasons ranging from calibration maneuvers to recoveries from instrument anomalies. Thus, the SW speed during these gaps have been removed to form sets of \{image, wind speed\}. 
%-----------------------
\subsection{Dataset partitioning and Cross Validation}
\begin{figure}
    \centering
\includegraphics[width=\linewidth]{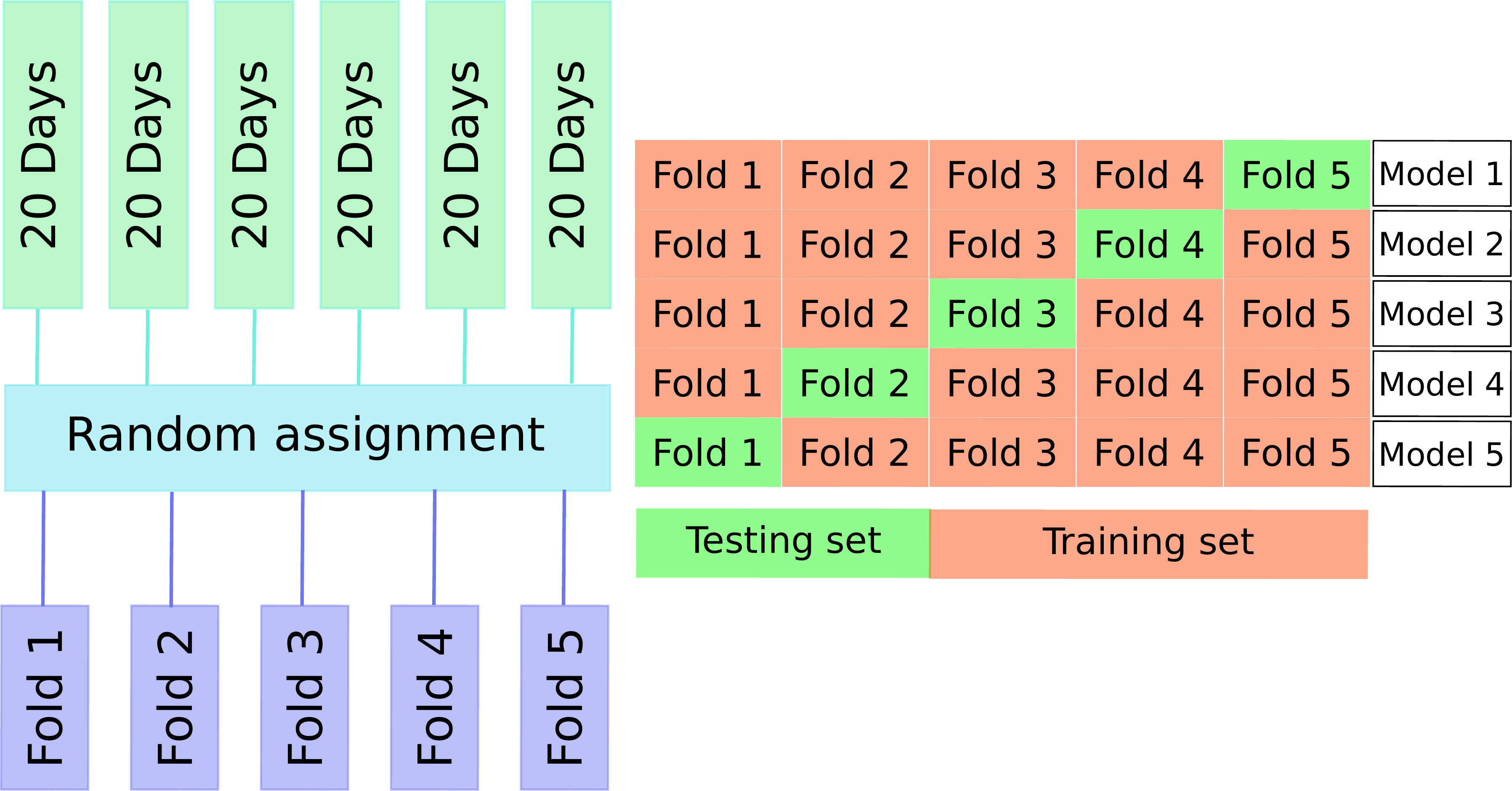}
    \caption{Training and test split for cross validation. First, the data are split into batches of 20 days each and each batch is randomly assigned to one of the cross-validation folds. Then, for one particular model (i.e., for one particular combination of history and delay), one of the folds is marked as the test set and the remaining as training sets. A circular permutation is performed till each fold is used as a test set. In our case, at the end of a training exercise, we will have 5 different variants of the particular model, from which we derive the mean and standard deviation of fitting metrics.}
    \label{fig:CVdataset}
\end{figure}

Data are available from 1 January 2011 to 9 December 2018. Given the presence of a background solar cycle and events in the Sun which might systematically bias our model to perform only for a particular phase of cycle, the whole dataset was partitioned into batches comprising 20 contiguous days of data. If during batch formation, there exists a single discontinuity in the batch, the data from the day prior to discontinuity to 20 days prior is sampled, and placed in the same place as the previous batch, thereby removing any data leak. If there exist multiple discontinuities (there are only 2 instances of such an event in either of the datasets), that particular window between the discontinuities is discarded. This results in 157 batches for 211~{\AA} and 158 batches for 193~{\AA} data (courtesy the one missing data, which resulted in a new batch). These batches were randomly sorted into 5 folds with equal probability, and these 5 folds were used to perform cross validation.  The dataset partitioning scheme is shown in Fig.~\ref{fig:CVdataset}.\\
In cross validation, if there are [1,N] folds of data, a cross validation set is constructed by holding the fold \emph{i} as the test set, and the remaining in training set. Such a construction is done for all folds of the batches. Our models are evaluated against this cross-validation dataset, thereby providing us with a mean value of the metric and a standard deviation. Henceforth, any standard deviation associated with the predictions are to be taken as evaluated on the cross-validation dataset.\\
The image data are centered using the mean pixel value of the training dataset per cross validation fold. The images are resized to $224\times224$ pixels using OpenCV's~\cite{opencv_library} default Linear Interpolation, and each image replicated into $3$ RGB channels. This was performed as our pre-trained network demands the input images to be of dimensions $224\times224\times3$, since terrestrial images generally have Red, Green and Blue as the color basis. These images are then finally used for training our network. The solar wind speed data are scaled between 0 and 1 using the training data statistics (max and min values) of each cross validation fold.
%---------------------------
\subsection{Control Hyperparameters}
\label{subsec:ControlParam}
\begin{figure}[ht!]
\includegraphics[width=\linewidth]{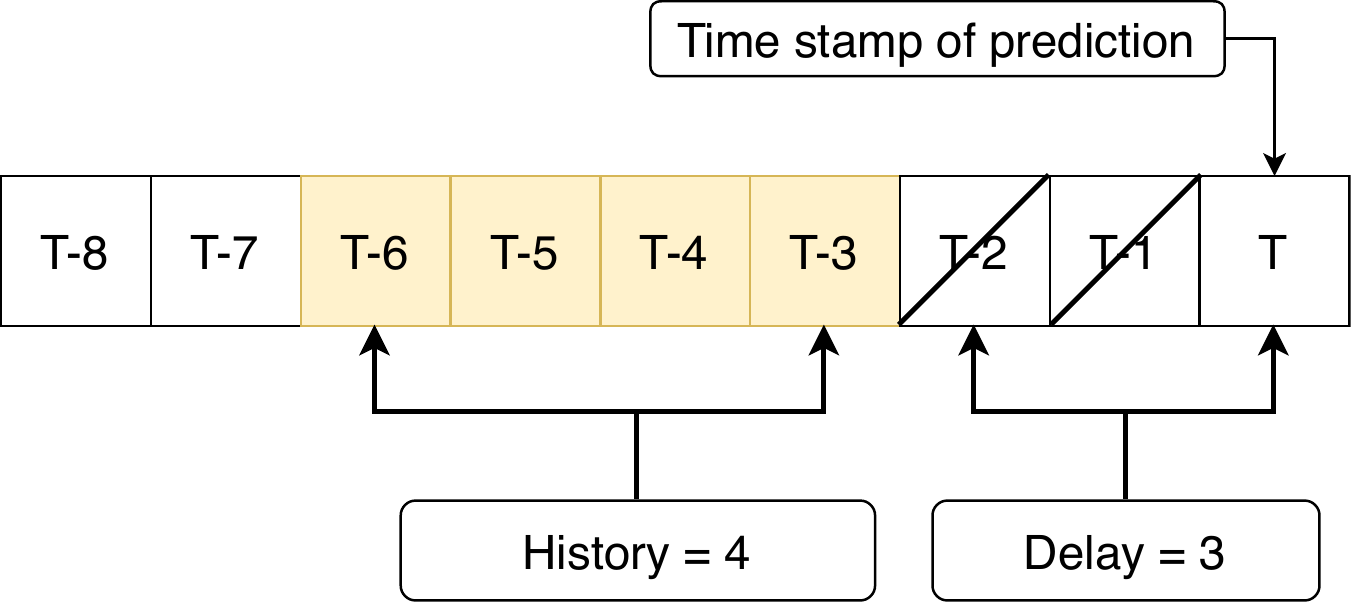}
\caption{Each variant of the WindNet model is trained to predict the SW speed on day $T$, using input data (SW and SDO/AIA) from days in the range [$T$-$H$-$D$+1,$T$-$D$], where $H$ and $D$ denote the history and delay control hyperparameters, respectively.}
\label{fig:hd}
\end{figure}

%To get a handle on the data, we
Hyperparameters are free parameters which give a handle in controlling the whole algorithm. We define two control hyperparameters: \textbf{history $H$} - number of days of input data required for one prediction and \textbf{delay $D$} - time from the latest input datapoint to the day of SW prediction.  For example, if the day of prediction is $T$, and data from $T$-3 to $T$-6 are used as input, our \emph{history} is defined as 4 and the \emph{delay} as 3. We have trained models with different combinations of delay ($D=$1 to 4) and history ($H=$1 to 4), resulting in 16 variants of the WindNet model.
The meaning of the two control hyperparameters are illustrated in Fig.~\ref{fig:hd}.
%---------------------
\subsection{Metrics for Comparison}
\label{sec:metric}
Quantitatively, a set of metrics need to be defined to unambiguously quantify if the fit is good or bad. We define three metrics to estimate the goodness of fit ($\hat{y}$ = Prediction, $y$ = Observation):
\begin{enumerate}
\item Mean square error ($\chi^2$ value):
	\begin{equation}
	\chi^2 = \frac{1}{N}\sum^{N}_{i}(\hat{y}_{i}-y_{i})^{2},
	\end{equation}
	where N is the no. of datapoints. However, we present the Root Mean Square Error (RMSE), defined as $\sqrt{\chi^2}$, in the units of $km/s$.
\item Reduced mean square error ($\chi^2_{red}$):
	\begin{equation}
	\chi_{red}^2 = \frac{1}{N}\sum^{N}_{i}\frac{(\hat{y}_{i}-y_{i})^{2}}{\sigma^{2}_i},
	\end{equation}
	where $\sigma_i$ denotes the standard deviation associated with each observation $y_i$. The standard deviation is computed over OMNI measurements for each day, and reported in the dataset.
\item Pearson correlation coefficient ($r$): The standard definition of correlation is
	\begin{equation}
	r = \frac{\sum^{N}_{i}(y_{i}-\bar{y})(\hat{y}_i-\hat{\bar{y}}) }{\sigma_{y}\sigma_{\hat{y}}},
	\end{equation}
	where $\bar{y}$ and $\hat{\bar{y}}$ represent the mean values and $\sigma_{y}$, $\sigma_{\hat{y}}$ represent the standard deviation of the dataset in consideration. To perform an average of correlation across all folds, we transform the data to Fischer's z-space, perform the averaging, and then transform back -- to prevent bias while performing average~\cite{Fischerzcorr}. The standard deviations are calculated in Fischer's z-space, and propagated back to correlation. 
\end{enumerate}
The three metrics defined have their own advantages and drawbacks.
\begin{enumerate}
\item The predicted data are scaled between \emph{0} and \emph{1}. Hence, even a large deviation, when squared, seems very small, if both the prediction and observation are $<1$. Thus, while $\chi^2$ is a good minimizing function for training, it fails to perform well as a metric for good fit.
\item To counter the above case, the $\chi_{red}^2$ metric is used. This takes into account the inherent error in each measurement, and scales the fit accordingly. A bad fit for a high error datapoint is acceptable, as the observation itself has high uncertainty, while a bad prediction on a low error datapoint is bad, since it serves as a much better point of comparison.
\item If the output were a naive mean value of the batch, the $\chi^2$ and $\chi_{red}^2$ would still be reasonable -- however, there would be no variance in the fit. Hence, the Pearson correlation $r$ is used to understand the trend captured by the fitted curve. The exact fit values may not match, but if the trend is captured, the model is fairly good according to this metric.
\end{enumerate}

Summarizing, Pearson $r$ captures the trend but ignores any scaling error. $\chi^2$ captures scaling errors, but doesn't perform well on scaled data $<1$. And $\chi_{red}^2$ captures the errors by weighing them vis-a-vis the variance of observed data. These three metrics are used for comparing the models - i.e, the $r$ value of our proposed model should be higher, and the $\chi^2$ and $\chi^2_{red}$ lesser than the benchmark models. Please note that errors (or spread) reported (for both the metrics, and activation evaluation) later on correspond to Standard Error (or uncertainty in the estimated mean), defined as $$S(x) := \frac{\sigma(x)}{\sqrt{N(x)}},$$ where $\sigma(x)$ is the standard deviation derived from the sample, and $N(x)$ is the number of samples in the set.\\
Model performance, while accounting for timing errors, is an importance marker for capturing the response to dynamic events in the SW. Thus, we also compare the performance of our models through its ability to capture High Speed Enhancements (HSE), as used in several texts~\cite{owensHSE,reissHSE,BuHSE}. We use the method as outlined in ~\citeA{jian2015validation} for finding out HSE. This is performed as:
\begin{enumerate}
\item Mark all time points which are more than 50 km/s faster than 1 day earlier.
%\item Eliminate any isolated single data points which are marked.
\item Group each contiguous block of marked points as a distinct high speed enhancement (HSE) and find the start and end time of each HSE.
\item For each HSE, find the minimum speed starting 2 days ahead of the HSE till the start of the HSE, and mark it as the minimum speed (Vmin) of the HSE; find the maximum speed starting from the beginning of the HSE through 1 day after the HSE and mark it as the maximum speed (Vmax) of the HSE. 
\item For each HSE, find the last time reaching Vmin and the first time reaching Vmax and mark them as the start and end time of an SIR.
\item For the regrouped SIRs, find the Vmin and Vmax for each SIR and mark the last time of highest speed gradient as the stream interface (SI), the boundary between slow and fast wind. Eliminate SIRs with the redundant SI time.
\item Reject any SIRs with Vmin faster than 500 km/s, or Vmax slower than 400 km, or speed increase less than 100 km/s.
\end{enumerate}
Each HSE present in the observation, and captured by the model is called a True Positive (TP), and those not captured by the model are called False Negative (FN). Spurious HSE predictions by model are called False Positives (FP). With these, we define the metric of comparison Threat Score (TS) as:
\begin{equation}
    TS=\frac{TP}{TP+FN+FP}.
    \label{eqn:TS}
\end{equation}
Threat score is a proxy for the accuracy of forecast of any model. A model which predicts all the HSE perfectly (while not predicting any spurious HSE) has a TS of 1 -- thus, lower the TS, worse the model. For every cross validation set per model, the HSEs are identified and TS calculated -- thereby giving us a mean TS and its uncertainty per model. Note that if the HSE (i.e the peak of the enhancement) occurs very near the boundary, it would be missed by the algorithm due to our data partitioning scheme. Such HSE are discarded by benchmarking the H=1,D=1 Persistence model to give a TS = 1.0. \\

This study does not account for the effect of ICMEs (Near-Earth Interplanetary Coronal Mass Ejections). There are 170 ICMEs reported within the time range considered in this study, affecting solar wind measurements in 336 days. In both model training and evaluation, we did not remove days for which there were ICMEs. The prediction of solar eruptions leading to CMEs and ICMEs is outside the scope of this study. Nevertheless, their occurrence impacts the solar wind measurements at L1. So for evaluation of the solar wind models in this paper, we decided to include even the days when ICMEs were present.

%-----------------------------------
\section{Modelling and methods} \label{sec:MandM}
\subsection{Benchmark Models}
We next describe various models taken as benchmarks for our proposed WindNet model. These benchmark models all operate as autoregressive models on the SW data only and do not use AIA images as input. The models (except 27-day persistence) are all corrected for the data gaps, thereby making the comparison reasonable.
\begin{itemize}
\item Naive mean value model.
\item N day and 27 day Persistence model.
\item Autoregression with XGBoost~\cite{Chen:2016:XST:2939672.2939785}.
\item Autoregression with Support Vector Machines (SVMs).
\end{itemize}
%In the following sections, we provide brief descriptions of the aforementioned models. %A more detailed introduction to the ML and DL methods is provided in the supporting information section. %\ref{sec:app1}.

\subsubsection{Autoregression using a `Mean value'}
One of the most basic benchmarks for any model is the comparison of the fit with a mean value model. This benchmark takes in the SW data and outputs the mean value of the whole batch. This model serves as the lowest benchmark that the proposed model should surpass, since untrained models output mean values.

\subsubsection{Persistence Model}
The second benchmark model is persistence.
The SW speed is fed in as input, and the same output is obtained. Such a model would show how long the data persists through time.

The N day persistence is calculated from $H+D-1$ days prior to prediction, to the day of prediction. As such, there is no individual dependence of the persistence model on $H$ or $D$ -- rather, the dependence is on the combined value, thereby having degeneracy. This model is primarily used for determining how far into the future our models consistently give a good prediction, given an observation today, or observations starting today.

We also benchmark our results against 27-day persistence for 1 Carrington rotation, as it has been shown to be a good benchmark model in~\citeA{Owens27DayPersistence}. The 27-day persistence model operated on the complete SW dataset (devoid of any gaps). %-- since a single day of gap will result in one entire Carrington rotation being neglected. The other models have cross validation, hence we could afford to resample the data for 20 days

\subsubsection{Autoregression using XGBoost}
The SW speed is autoregressed for different H and D using the XGboost algorithm~\cite{Chen:2016:XST:2939672.2939785}.
That is, the prediction $\hat{y}_{\rm T+1}$ is given as $\hat{y}_{\rm T+1} = f(\mathbf{x})$, where model input is $\mathbf{x} = (y_{\rm T-H-D+1}, y_{\rm T-H-D}, ..., y_{\rm T-D})$, and the function $f()$ comprises the gradient-boosted decision trees. The various parameters set for the algorithm are shown in Table.~\ref{tab:XGB}. The best model from the swept set of parameters is selected based on the lowest $\chi^2$ value.

\begin{table}
	\centering
	\caption{XGBoost parameter selection using grid search.}
	\label{tab:XGB}
	\begin{tabular}{l l}
		\toprule
		\textbf{Parameter} & \textbf{Value}\\
		\midrule
		eta & [0.001,0.01,0.1,0.8,0.9,1.0]\\
		seed & 0\\
		objective & reg:linear \\
		max\_depth & 200\\
		lambda & [50,10,5,1,0.5,0.05]\\
		\bottomrule
	\end{tabular}
\end{table}

\subsubsection{Autoregression using support vector machines (SVMs)}
SVM is also used as a benchmark for good fit, since it has more non-linearity than decision trees due to the presence of kernels. Three kernels are used for benchmarking - Radial Basis function, Linear, and Polynomial kernel of degree 5. We use the Scikit-learn~\cite{sklearn} implementation of SVM in this work. The parameters were selected by grid search using the $\chi^2$ value as the comparison metric. The best fitting parameters are shown in Table.~\ref{tab:SVM}.

\begin{table}
	\centering
	\caption{Support-vector regression-parameter selection.}
	\label{tab:SVM}
	\begin{tabular}{l l l}
		\toprule
		\textbf{Kernel} & \textbf{Parameter}& \textbf{Value}\\
		\midrule
		RBF & C & 1e+4 \\
		RBF & gamma & 0.001 \\
		Linear & C & 1e+4\\
		Polynomial & C & 1e+4\\
		Polynomial & degree	& 5 \\
		\bottomrule
	\end{tabular}
\end{table}
%-----
\subsection{Proposed SW model}
%\begin{figure}[ht!]
%\centering
%\includegraphics[width=\linewidth]{figures/inception.png}
%\caption{Inception module present in GoogleNet \cite{43022}.}
%\label{fig:incept}
%\end{figure}
\begin{figure}[ht!]
\includegraphics[width=\linewidth]{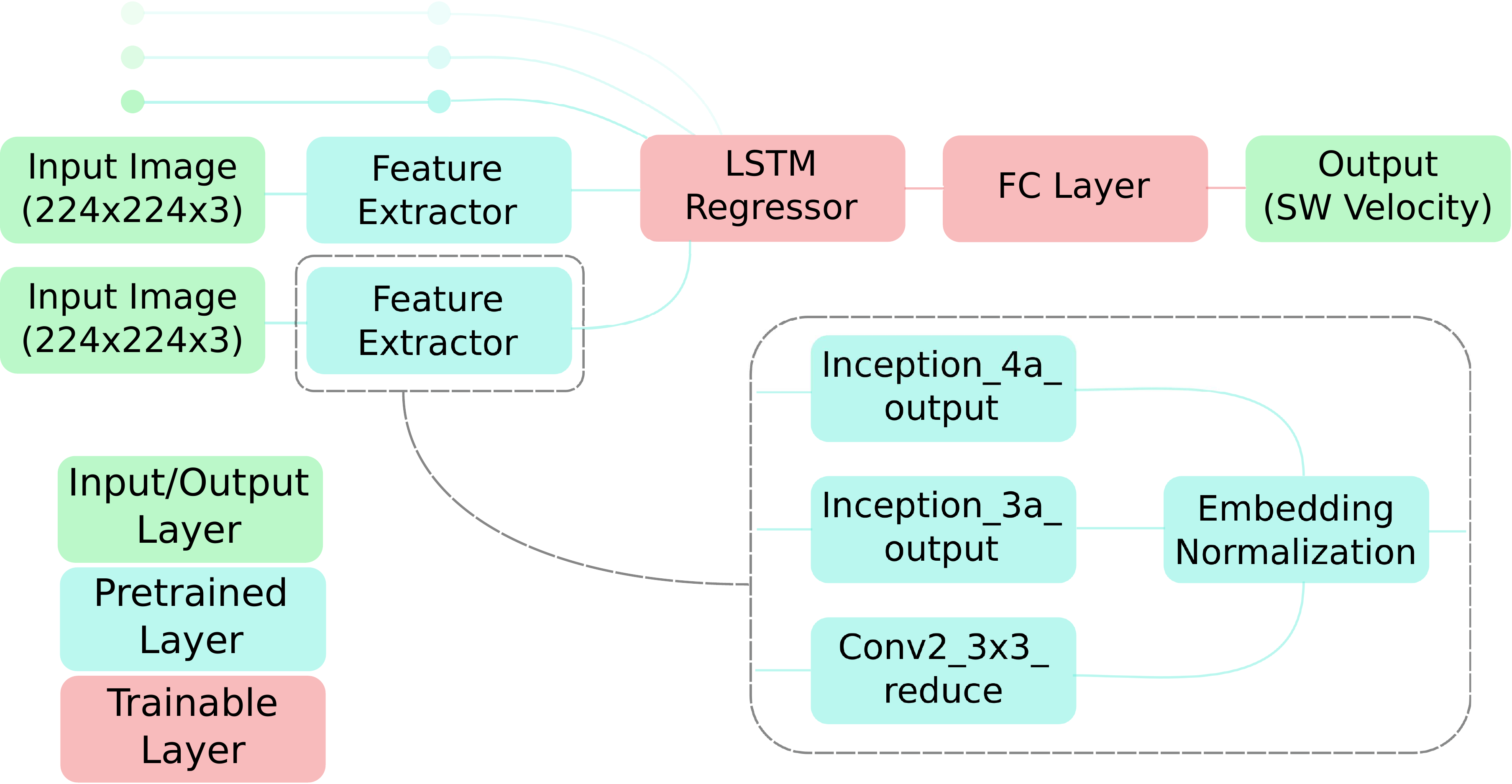}
\caption{WindNet architecture using GoogLeNet and an LSTM.}
\label{fig:custom}
\end{figure}

The methodology followed here is to define a feature extractor, which reduces the dimensionality of AIA images into a set of generic features, and a regressor which then takes these abstract features to regress against the solar wind speed. The proposed model \textbf{WindNet} is a deep learning model constructed using a ConvNet and a RNN. Here, a pretrained GoogLeNet~\cite{43022} model is used as a ConvNet feature extractor, and the obtained embeddings are fed into a variant of a RNN, called Long-Short Term Memory~\cite<LSTM;>[]{lstm} model (GoogLeNet weights were obtained from: http://www.deeplearningmodel.net/).

GoogLeNet is a ConvNet~\cite{43022} developed for the ImageNet~\cite{deng2009imagenet} competition. This competition provides a huge database of labelled images with the objective of classifying them into different categories. As mentioned in Sec.~\ref{sec:intro}, ConvNets work by detecting patterns at multiple scales in the input. This scale over which patterns are detected is given by the size of the convolution kernel. ConvNets generally have convolutions performed sequentially -- thus, at a given layer, the sensitivity is only to a particular scale. However, GoogLeNet, for the first time, introduces us to the concept of the \emph{Inception module}~\cite{43022}. Essentially, this module has, at each layer, convolutions using different kernel sizes done in parallel. Thus, it provides sensitivity at multiple scales at the same time. This has been shown to outperform other models on the ImageNet14 dataset~\cite{43022}. %The module is shown Fig.~\ref{fig:incept}
GoogLeNet has been trained on everyday objects. However, given the large volume of training data in ImageNet14, the initial layers of the network capture generic global features in the images~\cite{Goodfellow-et-al-2016}. As one goes deeper, the network captures features specific to the dataset -- which is not relevant to our dataset. Thus, we use this pretrained network and generate an embedding corresponding to the AIA data. This technique is known in the literature as Transfer learning~\cite{transfLearn}. We adopt a `multi-resolution approach' to generate the embeddings -- i.e, responses from layers at different depths are taken, normalized and concatenated. The embeddings are then fed to an LSTM for regression against the SW speed. GoogLeNet has its weights fixed, while the LSTM (and a fully connected layer at the end) are trained. We use a single LSTM cell in our work. The model is developed using the Tensorflow package for Python~\cite{tensorflow2015-whitepaper}, and has been summarized in Fig.~\ref{fig:custom}.
%While the feature extractor has been trained on everyday objects, given the large volume of training data in ImageNet14~\cite{deng2009imagenet} dataset, it is assumed the model has learned many types of filters and could potentially be used as a feature extractor, or an embedding generator for prediction using a technique called Transfer learning~\cite{transfLearn}. Generally, as one goes deeper into the network, local features are captured which do not appear to help in a transfer learning approach, and the use of the first few layers is preferred. However, a multi-resolution approach is adopted here, i.e., filter responses at different depths are taken, concatenated, and used for prediction. This set of embeddings is fed to an LSTM, and then to a fully connected layer for regression. The pretrained feature extractor network has its weights fixed, while the LSTM and regression layers are trained. The model is developed using the Tensorflow package for Python \cite{tensorflow2015-whitepaper}, and has been summarized in Fig.~\ref{fig:custom}.%\footnote{Model and tutorial available at: \url{https://github.com/Vishal-Upendran/WindNet}}. 

The training details for the algorithm are summarized in Table.~\ref{tab:SDO}.
\begin{table}
	\centering
	\caption{WindNet parameter selection}
	\label{tab:SDO}
	\begin{tabular}{l l}
		\toprule
		\textbf{Parameter} & \textbf{Value}\\
		\midrule
		Cost function & $\chi^2$($\hat{y}$,$y$)+$\chi^2_{red}$($\hat{y}$,$y$)\\
		Optimizer & Adam\\
		Learning rate & 5e-4\\
		Dropout for LSTM & 0.5\\
		L2 Norm coefficient& 1e-6\\
        No. of hidden units in one LSTM cell & 400 \\
        No. of iterations & 300 \\
        Feature length from GoogleNet & 832 \\
		\bottomrule
	\end{tabular}
\end{table}
%-----------
\subsection{Activation Visualization}
\label{sec:vis}
There exist techniques in the DL literature to visualize neurons in  hidden layers which are preferentially activated for a given input - this activation can be extrapolated back to the given input to understand which regions of the input data have large impact on the prediction. These methods rely primarily on the gradient of output w.r.t each input pixel, thereby providing an approximation of regions most responsible for an increase or decrease in the output. The methods, while not being perfect visualizers, are a window into the workings of the network. In this work, we use Grad-CAM~\cite{gradcam} maps as a visualization technique.
Grad-CAM, or Gradient Class Activation Maps, are maps generated by a pointwise multiplication of the average gradient per channel of output vis-a-vis a given convolution layer with the corresponding ConvNet layer activation. The obtained map is then passed through a Rectified Linear Unit (ReLU, namely $f(x) = {\rm max}([0,x])$) activation function to obtain the activation map. The maps are averaged across channels, and then scaled up to the dimensions of the input image for comparison.
This method produces activation maps of the model on the input data. These activation maps are subsequently used to generate a metric for the determination of influence of the CHs and ARs.
\subsubsection{Generating binary masks}
\label{sec:QGC}
A simple metric for understanding the influence of a particular set of features for a regression problem would be to look at the mean value of the activation on that particular set of features across all datapoints, and look for the variation of this mean value over days leading to prediction. Therefore it is of great importance to segment out the CHs and the ARs to generate binary maps.
\begin{figure}[ht!]
\centering
\includegraphics[width=\linewidth]{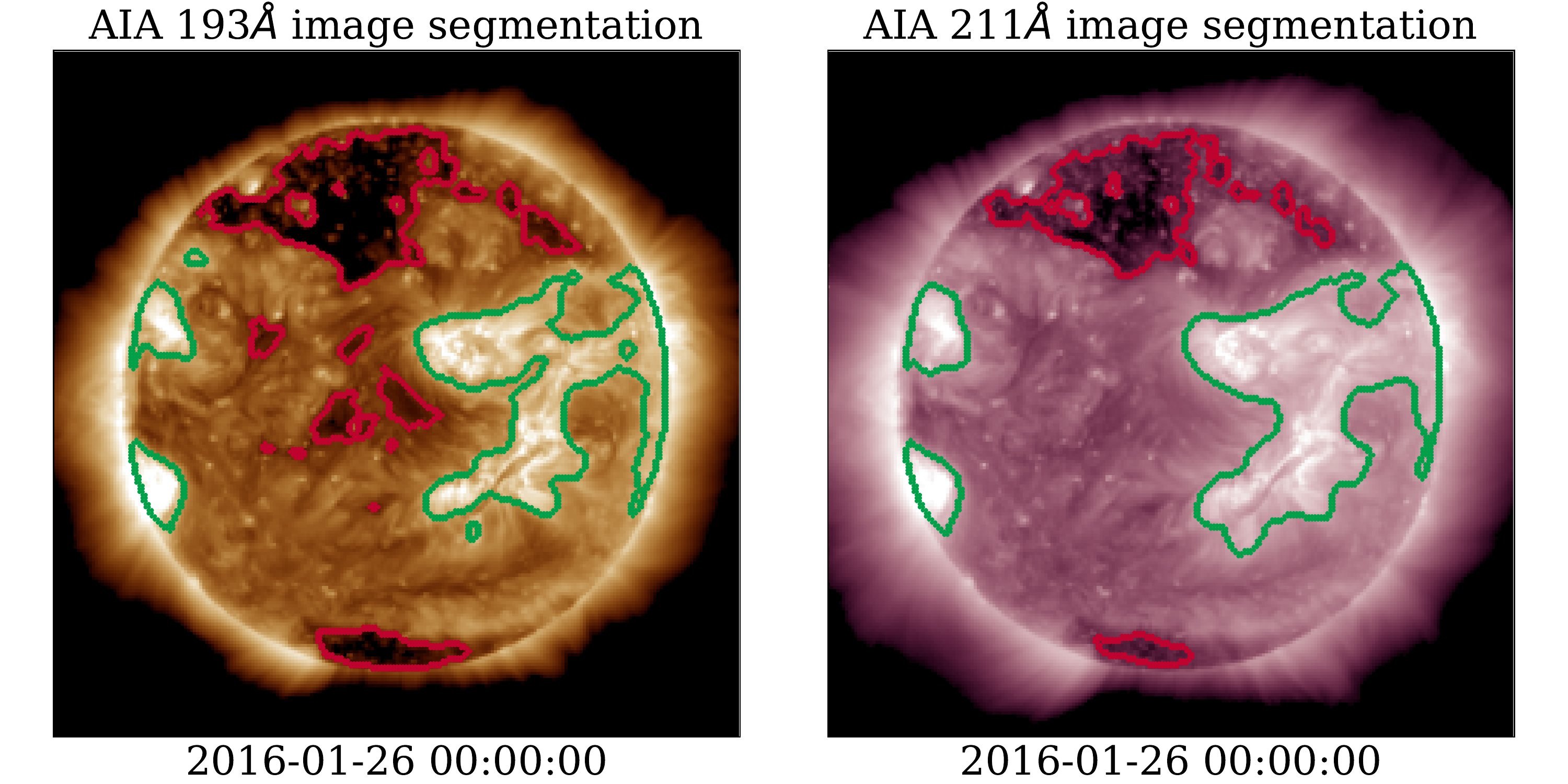}
\caption{A representative visualization of the segmentation map of 193~{\AA} channel (left) and 211~{\AA} channel (right) using classical computer vision algorithms. The overplotted green contours enclose AR, and the red contours CH. The segmentation maps are created separately for the AR and CH.}
\label{fig:segment}
\end{figure}
To obtain the CH segmentation map, we use Otsu thresholding~\cite{Otsu}. This thresholding assumes the presence of two distinct classes of pixels intensities, essentially Gaussian, and tries to find an intensity value which would maximize the inter-class variance (or alternatively, minimize the intra-class variance). We use stacked thresholding -- i.e, a preliminary threshold to segment out the approximate region of the coronal holes first, and then another threshold to segment out the coronal holes from this subset of the image.

The AR segmentation is far more non-trivial. Otsu thresholding picks out spurious areas as `active regions'. Hence, we apply a 5-class Gaussian Mixture Model ~\cite{sklearn} on the pixel intensities to segment out the ARs. The Gaussian with the highest mean is found to segment out the ARs well. A representative set of  segmentation maps overplotted on the EUV data is shown in Fig.~\ref{fig:segment}.

With these binary maps, we simply perform a pointwise multiplication of our activation values on a given image with its CH and AR map respectively, while also scaling by the total area of segmentation. The scaling by area of CH and AR is done to remove dependence on the absolute size of these regions, and obtain a normalized quantity. We then take the mean value over the image, and across all datasets to obtain a single scalar to quantify the activation at ARs and CHs, across the days of history for both fast and slow SW.  The activation plots are constructed for the training set (for better statistics), since the generalizability of the model is captured in its performance on the test set (or the cross-validation set).

%------------------------------------->
\section{Results \label{resultsec}}
\label{sec:result}
\subsection{Model benchmarking}
\begin{sidewaysfigure}
\centering
\includegraphics[width=50pc]{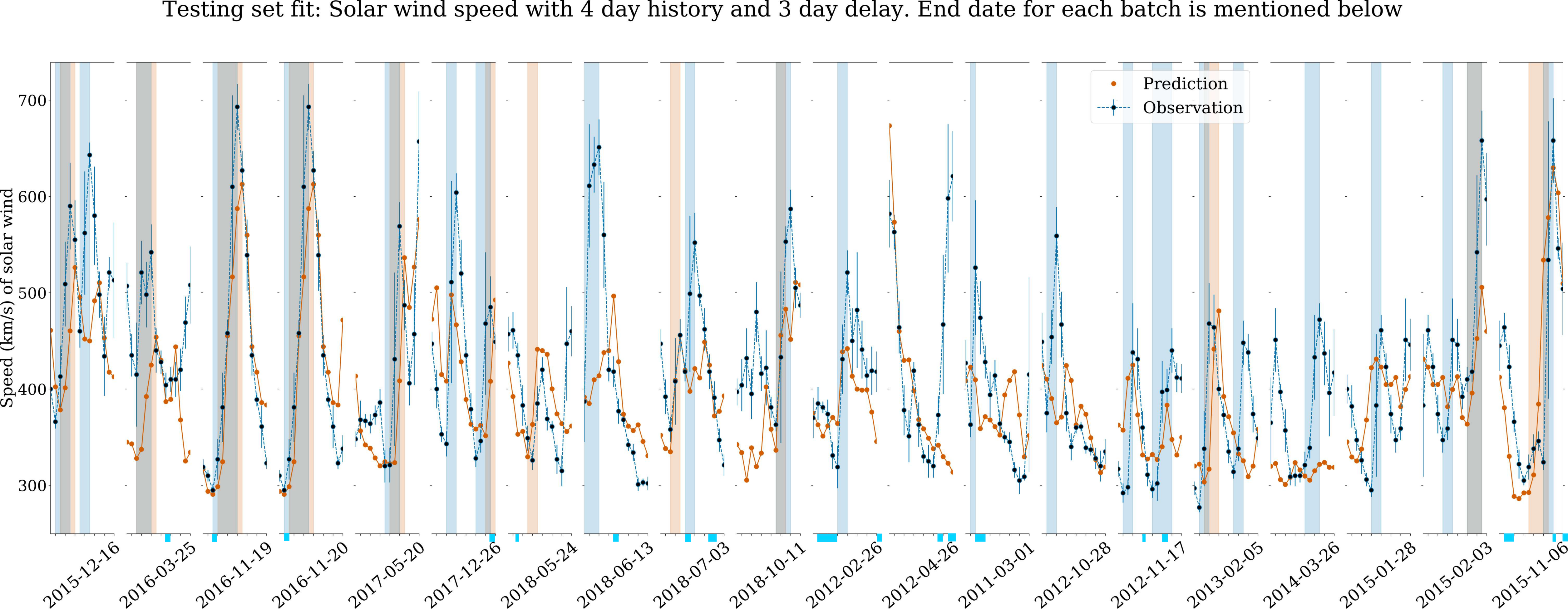}
\caption{Wind-speed prediction plot from one of the cross validation models, using 4 days of image data with 3 days of delay. On the x-axis, the ending date of each batch is shown. Since batches are randomly assigned to each cross-validation fold, the dates are not kept in order. The model has a correlation value of 0.61, RMSE of 76.4 km/s, and $\chi^2_{red}$ of 19.35. The errorbars are measurement errors of the wind-speed observations. The HSE are highlighted by their start and end times -- blue for the observed wind speed and red for the predicted wind speed. The blue bars below the plot indicate ICMEs. }
\label{fig:WindNet424}
\end{sidewaysfigure}
\begin{sidewaystable}
    \centering
    \caption{Correlation comparison of our model predictions with the Benchmark models. 27-day persistence gives a correlation of 0.456$\pm$0.02. Models which do not have a correlation value are given `--'. The p-values are all less than $10^{-7}$ for WindNet variants, and less than $10^{-2}$ for the benchmark models. }
    \label{tab:corrcomp}
    \begin{tabular}{ccccccccc}
    \hline 
(H,D) & WindNet 193 & WindNet 211 & XGBoost & Persistence & SVM Linear & SVM RBF & SVM Polynomial & Naive mean \\ 
\hline
(1,1) & 0.28 $\pm$ 0.03 & 0.34 $\pm$ 0.02 & 0.73 $\pm$ 0.01 & 0.76 $\pm$ 0.01 & 0.76 $\pm$ 0.01 & 0.76 $\pm$ 0.01 & 0.56 $\pm$ 0.01 &  -- \\ 
(1,2) & 0.37 $\pm$ 0.03 & 0.42 $\pm$ 0.03 & 0.36 $\pm$ 0.02 & 0.43 $\pm$ 0.02 & 0.43 $\pm$ 0.02 & 0.43 $\pm$ 0.02 & 0.34 $\pm$ 0.01 &  -- \\ 
(1,3) & 0.47 $\pm$ 0.01 & 0.48 $\pm$ 0.02 & 0.12 $\pm$ 0.02 & 0.19 $\pm$ 0.02 & 0.19 $\pm$ 0.03 & 0.19 $\pm$ 0.02 & 0.18 $\pm$ 0.03 &  -- \\ 
(1,4) & 0.46 $\pm$ 0.03 & 0.52 $\pm$ 0.04 & 0.02 $\pm$ 0.03 & 0.07 $\pm$ 0.03 & 0.08 $\pm$ 0.03 & 0.07 $\pm$ 0.03 & 0.07 $\pm$ 0.03 &  -- \\ 
(2,1) & 0.37 $\pm$ 0.05 & 0.42 $\pm$ 0.02 & 0.76 $\pm$ 0.01 & 0.43 $\pm$ 0.02 & 0.79 $\pm$ 0.01 & 0.79 $\pm$ 0.01 & 0.56 $\pm$ 0.01 &  -- \\ 
(2,2) & 0.47 $\pm$ 0.02 & 0.39 $\pm$ 0.06 & 0.40 $\pm$ 0.02 & 0.19 $\pm$ 0.02 & 0.47 $\pm$ 0.02 & 0.47 $\pm$ 0.02 & 0.34 $\pm$ 0.02 &  -- \\ 
(2,3) & 0.46 $\pm$ 0.03 & 0.53 $\pm$ 0.03 & 0.16 $\pm$ 0.02 & 0.07 $\pm$ 0.03 & 0.22 $\pm$ 0.02 & 0.21 $\pm$ 0.02 & 0.18 $\pm$ 0.03 &  -- \\ 
(2,4) & 0.51 $\pm$ 0.03 & 0.48 $\pm$ 0.03 & 0.01 $\pm$ 0.03 & 0.03 $\pm$ 0.03 & 0.08 $\pm$ 0.03 & 0.08 $\pm$ 0.03 & 0.05 $\pm$ 0.03 &  -- \\ 
(3,1) & 0.41 $\pm$ 0.04 & 0.51 $\pm$ 0.03 & 0.76 $\pm$ 0.01 & 0.19 $\pm$ 0.02 & 0.79 $\pm$ 0.01 & 0.79 $\pm$ 0.01 & 0.57 $\pm$ 0.01 &  -- \\ 
(3,2) & 0.46 $\pm$ 0.03 & 0.47 $\pm$ 0.02 & 0.39 $\pm$ 0.02 & 0.07 $\pm$ 0.03 & 0.47 $\pm$ 0.02 & 0.47 $\pm$ 0.02 & 0.34 $\pm$ 0.01 &  -- \\ 
(3,3) & 0.47 $\pm$ 0.03 & 0.53 $\pm$ 0.03 & 0.15 $\pm$ 0.02 & 0.03 $\pm$ 0.03 & 0.22 $\pm$ 0.02 & 0.22 $\pm$ 0.02 & 0.16 $\pm$ 0.02 &  -- \\ 
(3,4) & 0.46 $\pm$ 0.03 & 0.54 $\pm$ 0.03 & 0.03 $\pm$ 0.04 & 0.01 $\pm$ 0.03 & 0.08 $\pm$ 0.03 & 0.09 $\pm$ 0.03 & 0.05 $\pm$ 0.03 &  -- \\ 
(4,1) & 0.47 $\pm$ 0.04 & 0.54 $\pm$ 0.03 & 0.75 $\pm$ 0.01 & 0.07 $\pm$ 0.03 & 0.79 $\pm$ 0.01 & 0.79 $\pm$ 0.01 & 0.57 $\pm$ 0.01 &  -- \\ 
(4,2) & 0.48 $\pm$ 0.03 & 0.52 $\pm$ 0.02 & 0.38 $\pm$ 0.01 & 0.03 $\pm$ 0.03 & 0.47 $\pm$ 0.02 & 0.47 $\pm$ 0.02 & 0.34 $\pm$ 0.01 &  -- \\ 
(4,3) & 0.45 $\pm$ 0.04 & 0.55 $\pm$ 0.03 & 0.16 $\pm$ 0.02 & 0.01 $\pm$ 0.03 & 0.22 $\pm$ 0.02 & 0.22 $\pm$ 0.01 & 0.15 $\pm$ 0.03 &  -- \\ 
(4,4) & 0.48 $\pm$ 0.04 & 0.50 $\pm$ 0.03 & 0.04 $\pm$ 0.04 & -0.02 $\pm$ 0.03 & 0.09 $\pm$ 0.03 & 0.09 $\pm$ 0.03 & 0.06 $\pm$ 0.04 &  -- \\ 
\hline 
    \end{tabular}
\end{sidewaystable}
%---------
\begin{sidewaystable}
    \centering
    \caption{RMSE comparison of our model predictions with the Benchmark models. 27-day persistence gives a RMSE of 93.14$\pm$4.43.}
    \label{tab:msecomp}
    \begin{tabular}{ccccccccc}
    \hline
    (H,D) &  WindNet 193 & WindNet 211 & XGBoost & Persistence & SVM Linear & SVM RBF & SVM Polynomial & Naive mean\\  
    \hline 
(1,1) & 97.01 $\pm$ 4.02 & 96.64 $\pm$ 4.27 & 60.31 $\pm$ 0.62 & 62.02 $\pm$ 1.05 & 57.68 $\pm$ 0.72 & 57.68 $\pm$ 0.73 & 74.17 $\pm$ 1.32 & 88.05 $\pm$ 2.08 \\ 
(1,2) & 92.13 $\pm$ 2.88 & 89.45 $\pm$ 2.68 & 83.47 $\pm$ 1.00 & 95.35 $\pm$ 1.60 & 80.60 $\pm$ 1.28 & 80.55 $\pm$ 1.28 & 83.74 $\pm$ 1.75 & 87.77 $\pm$ 2.20 \\ 
(1,3) & 83.70 $\pm$ 1.77 & 87.34 $\pm$ 3.85 & 90.33 $\pm$ 1.30 & 113.81 $\pm$ 2.39 & 87.77 $\pm$ 1.74 & 87.78 $\pm$ 1.75 & 87.97 $\pm$ 1.91 & 88.04 $\pm$ 2.42 \\ 
(1,4) & 84.33 $\pm$ 2.31 & 85.94 $\pm$ 4.67 & 92.14 $\pm$ 1.73 & 122.20 $\pm$ 3.13 & 89.06 $\pm$ 1.98 & 89.07 $\pm$ 1.97 & 88.91 $\pm$ 1.94 & 88.29 $\pm$ 2.46 \\ 
(2,1) & 96.31 $\pm$ 4.87 & 91.12 $\pm$ 2.30 & 57.87 $\pm$ 0.65 & 95.35 $\pm$ 1.60 & 54.27 $\pm$ 0.93 & 54.19 $\pm$ 0.92 & 74.64 $\pm$ 1.80 & 87.77 $\pm$ 2.20 \\ 
(2,2) & 90.80 $\pm$ 2.85 & 102.85 $\pm$ 9.00 & 83.48 $\pm$ 0.83 & 113.81 $\pm$ 2.39 & 78.94 $\pm$ 1.58 & 78.98 $\pm$ 1.57 & 84.36 $\pm$ 1.90 & 88.04 $\pm$ 2.42 \\ 
(2,3) & 86.21 $\pm$ 2.12 & 83.38 $\pm$ 2.78 & 91.86 $\pm$ 1.19 & 122.20 $\pm$ 3.13 & 87.09 $\pm$ 1.73 & 87.17 $\pm$ 1.70 & 87.91 $\pm$ 1.85 & 88.29 $\pm$ 2.46 \\ 
(2,4) & 86.24 $\pm$ 2.63 & 86.53 $\pm$ 2.27 & 93.11 $\pm$ 1.14 & 125.16 $\pm$ 3.12 & 88.68 $\pm$ 1.95 & 88.72 $\pm$ 1.97 & 88.77 $\pm$ 1.88 & 88.86 $\pm$ 2.46 \\ 
(3,1) & 93.35 $\pm$ 4.33 & 82.60 $\pm$ 1.75 & 57.80 $\pm$ 0.80 & 113.81 $\pm$ 2.39 & 54.40 $\pm$ 1.01 & 54.34 $\pm$ 1.00 & 73.84 $\pm$ 1.70 & 88.04 $\pm$ 2.42 \\ 
(3,2) & 88.17 $\pm$ 1.81 & 85.46 $\pm$ 2.63 & 84.10 $\pm$ 0.86 & 122.20 $\pm$ 3.13 & 78.59 $\pm$ 1.67 & 78.55 $\pm$ 1.63 & 83.91 $\pm$ 1.85 & 88.29 $\pm$ 2.46 \\ 
(3,3) & 87.04 $\pm$ 1.25 & 83.97 $\pm$ 3.04 & 91.58 $\pm$ 1.05 & 125.16 $\pm$ 3.12 & 86.68 $\pm$ 1.76 & 86.75 $\pm$ 1.73 & 87.91 $\pm$ 1.79 & 88.86 $\pm$ 2.46 \\ 
(3,4) & 87.21 $\pm$ 2.17 & 81.21 $\pm$ 1.86 & 92.72 $\pm$ 1.28 & 126.79 $\pm$ 2.92 & 88.72 $\pm$ 1.75 & 88.62 $\pm$ 1.80 & 88.96 $\pm$ 1.67 & 89.14 $\pm$ 2.41 \\ 
(4,1) & 84.19 $\pm$ 2.83 & 80.27 $\pm$ 2.07 & 59.14 $\pm$ 0.82 & 122.20 $\pm$ 3.13 & 54.52 $\pm$ 1.03 & 54.48 $\pm$ 1.04 & 74.28 $\pm$ 2.07 & 88.29 $\pm$ 2.46 \\ 
(4,2) & 86.42 $\pm$ 1.98 & 83.06 $\pm$ 2.51 & 83.78 $\pm$ 0.74 & 125.16 $\pm$ 3.12 & 78.47 $\pm$ 1.81 & 78.45 $\pm$ 1.78 & 84.16 $\pm$ 1.88 & 88.86 $\pm$ 2.46 \\ 
(4,3) & 88.32 $\pm$ 1.93 & 80.28 $\pm$ 3.05 & 91.00 $\pm$ 1.25 & 126.79 $\pm$ 2.92 & 86.81 $\pm$ 1.59 & 86.82 $\pm$ 1.63 & 88.25 $\pm$ 1.68 & 89.14 $\pm$ 2.41 \\ 
(4,4) & 82.93 $\pm$ 1.72 & 85.34 $\pm$ 3.10 & 92.34 $\pm$ 1.34 & 128.23 $\pm$ 2.96 & 88.87 $\pm$ 1.46 & 88.78 $\pm$ 1.58 & 89.41 $\pm$ 1.40 & 89.43 $\pm$ 2.29 \\ 
\hline 
    \end{tabular}
\end{sidewaystable}
%---------
\begin{sidewaystable}
    \centering
    \caption{$\chi^2_{red}$ comparison of our model predictions with the Benchmark models. 27-day persistence gives a $\chi^2_{red}$ of 51.69$\pm$9.14.}
    \label{tab:redmsecomp}
    \begin{tabular}{ccccccccc}
    \hline
    (H,D)& WindNet 193 & WindNet 211 & XGBoost & Persistence & SVM Linear & SVM RBF & SVM Polynomial & Naive mean \\  
    \hline 
(1,1) & 33.38 $\pm$ 4.10 & 29.63 $\pm$ 1.36 & 23.18 $\pm$ 0.82 & 24.50 $\pm$ 0.73 & 21.20 $\pm$ 0.77 & 21.20 $\pm$ 0.77 & 35.17 $\pm$ 1.91 & 41.98 $\pm$ 3.71 \\ 
(1,2) & 28.43 $\pm$ 1.98 & 30.80 $\pm$ 2.05 & 44.10 $\pm$ 1.75 & 57.63 $\pm$ 1.76 & 41.13 $\pm$ 1.77 & 41.09 $\pm$ 1.79 & 44.54 $\pm$ 2.65 & 39.16 $\pm$ 3.92 \\ 
(1,3) & 25.82 $\pm$ 1.39 & 26.27 $\pm$ 1.97 & 51.67 $\pm$ 2.08 & 81.61 $\pm$ 2.74 & 48.79 $\pm$ 2.17 & 48.80 $\pm$ 2.20 & 49.09 $\pm$ 2.57 & 40.34 $\pm$ 3.25 \\ 
(1,4) & 26.83 $\pm$ 2.31 & 26.01 $\pm$ 2.47 & 54.21 $\pm$ 2.55 & 95.19 $\pm$ 4.25 & 50.63 $\pm$ 2.40 & 50.64 $\pm$ 2.39 & 50.48 $\pm$ 2.48 & 44.99 $\pm$ 3.65 \\ 
(2,1) & 48.07 $\pm$ 11.96 & 43.75 $\pm$ 7.58 & 21.17 $\pm$ 0.67 & 57.63 $\pm$ 1.76 & 18.63 $\pm$ 0.77 & 18.59 $\pm$ 0.77 & 35.44 $\pm$ 2.38 & 39.16 $\pm$ 3.92 \\ 
(2,2) & 47.32 $\pm$ 8.43 & 90.87 $\pm$ 38.04 & 44.05 $\pm$ 1.05 & 81.61 $\pm$ 2.74 & 39.41 $\pm$ 1.49 & 39.45 $\pm$ 1.50 & 45.15 $\pm$ 2.44 & 40.34 $\pm$ 3.25 \\ 
(2,3) & 30.09 $\pm$ 3.42 & 31.60 $\pm$ 3.84 & 53.88 $\pm$ 2.23 & 95.19 $\pm$ 4.25 & 48.38 $\pm$ 2.00 & 48.46 $\pm$ 1.99 & 49.37 $\pm$ 2.45 & 44.99 $\pm$ 3.65 \\ 
(2,4) & 31.41 $\pm$ 3.97 & 37.90 $\pm$ 4.07 & 54.86 $\pm$ 1.92 & 99.64 $\pm$ 5.18 & 49.71 $\pm$ 1.90 & 49.76 $\pm$ 1.91 & 49.83 $\pm$ 1.96 & 49.46 $\pm$ 3.69 \\ 
(3,1) & 47.56 $\pm$ 6.01 & 29.16 $\pm$ 2.87 & 21.13 $\pm$ 0.65 & 81.61 $\pm$ 2.74 & 18.71 $\pm$ 0.63 & 18.67 $\pm$ 0.62 & 34.60 $\pm$ 1.90 & 40.34 $\pm$ 3.25 \\ 
(3,2) & 33.51 $\pm$ 3.20 & 36.66 $\pm$ 3.87 & 45.06 $\pm$ 1.14 & 95.19 $\pm$ 4.25 & 39.34 $\pm$ 1.41 & 39.31 $\pm$ 1.41 & 44.97 $\pm$ 2.29 & 44.99 $\pm$ 3.65 \\ 
(3,3) & 43.29 $\pm$ 4.10 & 37.54 $\pm$ 7.33 & 53.09 $\pm$ 1.86 & 99.64 $\pm$ 5.18 & 47.47 $\pm$ 1.55 & 47.56 $\pm$ 1.57 & 48.89 $\pm$ 1.92 & 49.46 $\pm$ 3.69 \\ 
(3,4) & 42.35 $\pm$ 5.76 & 31.52 $\pm$ 5.16 & 53.40 $\pm$ 2.06 & 101.18 $\pm$ 5.53 & 48.83 $\pm$ 1.79 & 48.73 $\pm$ 1.85 & 49.12 $\pm$ 1.90 & 50.96 $\pm$ 5.72 \\ 
(4,1) & 31.16 $\pm$ 1.76 & 31.18 $\pm$ 3.36 & 22.25 $\pm$ 0.40 & 95.19 $\pm$ 4.25 & 18.92 $\pm$ 0.54 & 18.89 $\pm$ 0.54 & 35.33 $\pm$ 2.23 & 44.99 $\pm$ 3.65 \\ 
(4,2) & 27.99 $\pm$ 3.35 & 29.59 $\pm$ 3.94 & 44.34 $\pm$ 0.80 & 99.64 $\pm$ 5.18 & 38.87 $\pm$ 1.24 & 38.86 $\pm$ 1.25 & 44.81 $\pm$ 1.93 & 49.46 $\pm$ 3.69 \\ 
(4,3) & 36.98 $\pm$ 3.82 & 26.83 $\pm$ 2.20 & 51.47 $\pm$ 2.06 & 101.18 $\pm$ 5.53 & 46.73 $\pm$ 1.53 & 46.76 $\pm$ 1.62 & 48.36 $\pm$ 2.02 & 50.96 $\pm$ 5.72 \\ 
(4,4) & 31.90 $\pm$ 3.38 & 32.58 $\pm$ 4.83 & 52.66 $\pm$ 2.21 & 103.16 $\pm$ 5.66 & 48.70 $\pm$ 1.81 & 48.62 $\pm$ 1.94 & 49.35 $\pm$ 2.06 & 52.29 $\pm$ 4.11 \\ 
\hline 
    \end{tabular}
\end{sidewaystable}
%---------
\begin{sidewaystable}
    \centering
    \caption{HSE Threat Score comparison. The 27-day persistence model gives a TS of $0.506\pm0.029$. Cases with TS $0.0$, imply a value less than $1e-3$.}
    \label{tab:hsethreat}
    \begin{tabular}{ccccccccc}
    \hline
    (H,D) & WindNet 193 & WindNet 211 & XGBoost & Persistence & SVM Linear & SVM RBF & SVM Polynomial & Naive mean \\  
    \hline 
(1,1) & 0.081$\pm$0.023 & 0.112$\pm$0.040 & 0.776$\pm$0.037 & 1.000$\pm$0.000 & 0.748$\pm$0.038 & 0.748$\pm$0.038 & 0.263$\pm$0.026 & 0.0 \\ 
(1,2) & 0.042$\pm$0.022 & 0.150$\pm$0.042 & 0.329$\pm$0.012 & 0.858$\pm$0.037 & 0.288$\pm$0.027 & 0.271$\pm$0.028 & 0.162$\pm$0.035 & 0.0 \\ 
(1,3) & 0.167$\pm$0.008 & 0.140$\pm$0.041 & 0.061$\pm$0.015 & 0.351$\pm$0.021 & 0.0 & 0.0 & 0.029$\pm$0.013 & 0.0 \\ 
(1,4) & 0.212$\pm$0.042 & 0.206$\pm$0.047 & 0.036$\pm$0.018 & 0.199$\pm$0.024 & 0.0 & 0.0 & 0.0 & 0.0 \\ 
(2,1) & 0.203$\pm$0.064 & 0.227$\pm$0.029 & 0.711$\pm$0.036 & 0.858$\pm$0.037 & 0.850$\pm$0.022 & 0.845$\pm$0.021 & 0.292$\pm$0.027 & 0.0 \\ 
(2,2) & 0.293$\pm$0.022 & 0.297$\pm$0.040 & 0.423$\pm$0.096 & 0.351$\pm$0.021 & 0.461$\pm$0.017 & 0.449$\pm$0.017 & 0.148$\pm$0.033 & 0.0 \\ 
(2,3) & 0.225$\pm$0.036 & 0.198$\pm$0.037 & 0.030$\pm$0.027 & 0.199$\pm$0.024 & 0.0 & 0.0 & 0.020$\pm$0.011 & 0.0 \\ 
(2,4) & 0.239$\pm$0.045 & 0.282$\pm$0.044 & 0.043$\pm$0.038 & 0.215$\pm$0.022 & 0.0 & 0.0 & 0.0 & 0.0 \\ 
(3,1) & 0.310$\pm$0.051 & 0.259$\pm$0.031 & 0.753$\pm$0.024 & 0.351$\pm$0.021 & 0.850$\pm$0.026 & 0.844$\pm$0.026 & 0.323$\pm$0.026 & 0.0 \\ 
(3,2) & 0.237$\pm$0.048 & 0.292$\pm$0.029 & 0.472$\pm$0.107 & 0.199$\pm$0.024 & 0.426$\pm$0.018 & 0.408$\pm$0.024 & 0.113$\pm$0.029 & 0.0 \\ 
(3,3) & 0.328$\pm$0.038 & 0.287$\pm$0.017 & 0.116$\pm$0.047 & 0.215$\pm$0.022 & 0.0 & 0.0 & 0.0 & 0.0 \\ 
(3,4) & 0.357$\pm$0.031 & 0.294$\pm$0.026 & 0.024$\pm$0.022 & 0.236$\pm$0.026 & 0.0 & 0.0 & 0.0 & 0.0 \\ 
(4,1) & 0.286$\pm$0.037 & 0.309$\pm$0.027 & 0.737$\pm$0.034 & 0.199$\pm$0.024 & 0.849$\pm$0.032 & 0.845$\pm$0.034 & 0.292$\pm$0.031 & 0.0 \\ 
(4,2) & 0.298$\pm$0.040 & 0.200$\pm$0.039 & 0.428$\pm$0.083 & 0.215$\pm$0.022 & 0.431$\pm$0.024 & 0.428$\pm$0.025 & 0.157$\pm$0.037 & 0.0 \\ 
(4,3) & 0.289$\pm$0.070 & 0.200$\pm$0.056 & 0.115$\pm$0.044 & 0.236$\pm$0.026 & 0.0 & 0.015$\pm$0.009 & 0.011$\pm$0.010 & 0.0 \\ 
(4,4) & 0.251$\pm$0.049 & 0.314$\pm$0.080 & 0.035$\pm$0.022 & 0.307$\pm$0.033 & 0.0 & 0.0 & 0.007$\pm$0.006 & 0.0 \\ 
\hline 
    \end{tabular}
\end{sidewaystable}
%-----------
From Table.~\ref{tab:corrcomp} through Table.~\ref{tab:hsethreat}, we have summarized the performance of WindNet, as well as the benchmark autoregressive models for the metrics defined -- Correlation ($r$), RMSE, $\chi^2_{red}$ and TS respectively . We see that  WindNet outperforms the benchmarks over combinations where delay is generally more than 1 - i.e, where the autoregressive models do not have the immediately preceding solar wind speed available. In fact, for larger delays and histories, WindNet shows consistent performance, while other models fail to perform a reasonable prediction. The best performance of WindNet is for a history-delay combination of $(4,3)$, wherein the correlation is $\approx$ 0.55, and the spread is 0.03 -- this is for 211~{\AA}. Similarly, the best fit using 193~{\AA} data occurs for a combination of $(2,4)$, with a correlation of 0.51, with a spread of 0.03.

The Naive mean model has no variance, so there would be no correlation associated with it -- however, it is presented for the sake of completeness. Autoregressive SVM using an RBF kernel seems to perform better given the SW speed closer to the day of prediction, but falter as more delay is induced. The linear SVM performs as well as the non-linear RBF kernel, but the polynomial kernel fails to get a good fit. The 27-day persistence is a set of just 5 models -- thus, this performance is stated in the caption of the respective Tables. % The correlation plots using the Naive mean predictor and various SVM kernels are shown in Fig.~\ref{fig:svmcorr}, while those for Persistence and XGBoost are shown in Fig.~\ref{fig:perXG}. 
%-----
\subsection{WindNet prediction}
In this section we investigate the variation in prediction for our WindNet models. The model with highest correlation, as mentioned previously, is for a history of 4 and delay of 3 for 211~{\AA}. As can be seen in the Table.~\ref{tab:corrcomp}, there seems to be a subtle trend of an increase in correlation with history for a given delay for short delays. The predictions for models with delay smaller than history seem mostly consistent within the error bars. For the 193 {\AA} model in Table.~\ref{tab:corrcomp}, it can be seen that an increase in delay for a given history results in almost a consistent prediction correlation for high history models (again, within the errorbars -- though the mean values do not seem to follow an ordered trend), except in the case of 1 day history, where the correlation increases. This trend of increase in delay for a given history is largely followed in the 211~{\AA} data, though the $4$-day history seems to be the most consistent in this case within the errors, and the best performing. In general, the expectation would be an increase in correlation with increasing history, and some form of variation due to an increase in delay. The variation in performance with history for small delays is fairly consistent between both 193~{\AA} and 211~{\AA} with only the actual correlation values being different -- however, larger delay models do not have the same variation in performance for 193~{\AA} and 211~{\AA}. 211~{\AA}, in fact seems to be a better channel for SW prediction, since the corresponding models have higher correlation means and smaller standard deviations. Short-delay and short-history models (for example, 1 day history and delay) do not perform as well as models with larger history and delay (for example,4 day history and 3 day delay), since the solar wind is yet to arrive at L1. 193~{\AA} data shows a peak in correlation at 2 day history and 4 day delay. The 211~{\AA} data shows a similar peak at 4 day history and 3 day delay. 

A summary of RMSE is shown in Table.~\ref{tab:msecomp}, and a similar summary of $\chi^2_{red}$ is shown in Table.~\ref{tab:redmsecomp}. The TS is tabulated in Table.~\ref{tab:hsethreat}. The TS table shows that our proposed model has a maximum of 0.357$\pm$0.03. The low TS may be explained better by a careful observation of Fig.~\ref{fig:WindNet424}. This is a plot of one of the cross validation models using 211~{\AA}, having the highest correlation, with 4 day history and 3 day delay of data. With 10 TP, 2 FP, and 13 FN, the model is seen to have a TS of 0.4, a correlation with observation of 0.61, RMSE of 76.4 km/s and $\chi^2_{red}$ of 19.35. Here, we see that there are many more HSE present in the observed wind speed, which seem to be missing from the prediction. However, upon careful observation,it may be seen that many of the observed HSE do correspond to an enhancement in the wind speed of the predictor -- either at the exact time step, or with a lag/lead of 3 to 4 days. However, the predicted values do not show the drastic enhancement more than the prescribed thresholds. Thus, these events are not marked as HSE.
%    \item The $11^{th}$ observed HSE corresponds to an ICME. Since our model is not trained to account for ICMEs, it is not predicted. 
%    \item Many events are certainly missed by the algorithm if the HSE occur near the edge -- in such cases, the time window of matching gets reduced due to the start or the end point.
%\end{itemize}
The WindNet performance on the error metrics, though, largely complements the correlation performance, and show WindNet has better performance than the benchmark models for delays larger than 1 day in most cases.
%--------------------------
\subsection{Activation visualization}
\begin{figure}[!ht]
\includegraphics[width=\linewidth]{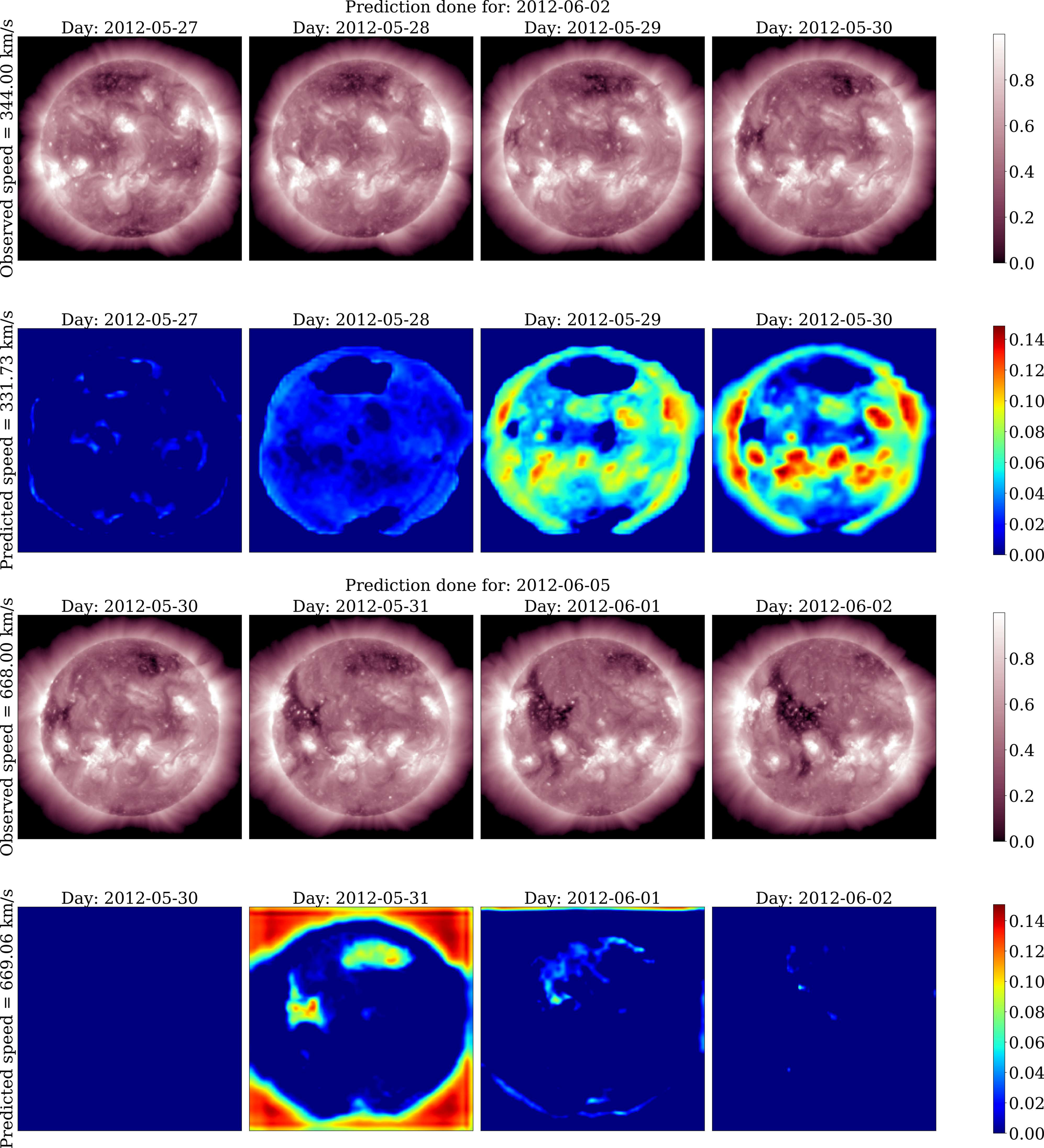}
\caption{GC activation maps for a fast (top) and slow wind (above) prediction using 211~{\AA} data, with the colormap corresponding to each row given on the right. The activation maps and the images have been rescaled between 0 and 1 row-wise for ease of comparison. For the fast wind prediction, note how the maximum activation occurs at the CH, 3 to 4 days prior to prediction, which seems to match with the correlations obtained in the literature ~\cite{CorHole_Vrsnak}. The slow wind, on the other hand activates the AR closer to the prediction, with no activation at the CH. }
\label{fig:GC2}
\end{figure}
\begin{figure}[!ht]
\includegraphics[width=\linewidth]{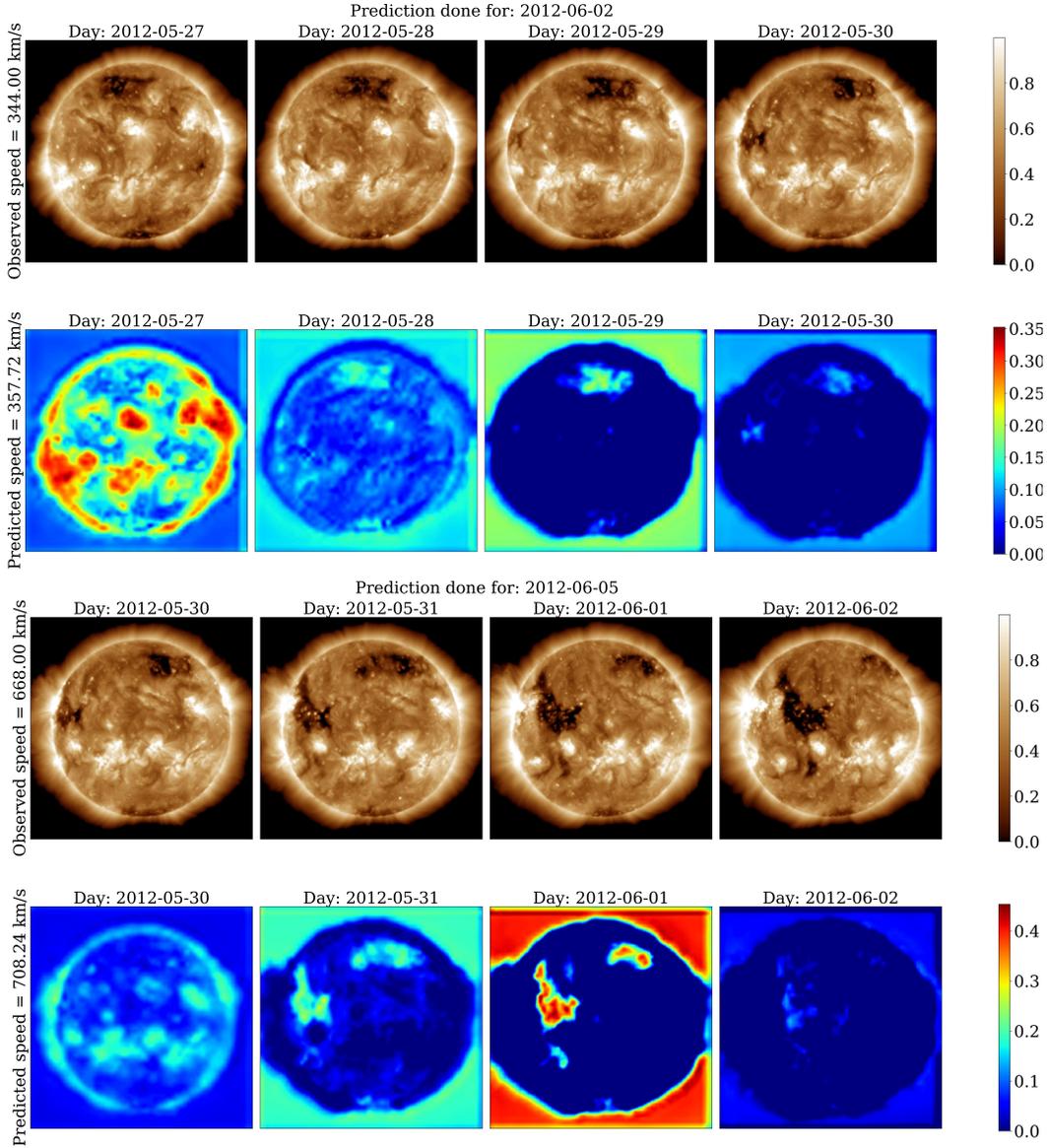}
\caption{GC activation maps for a fast (top) and slow wind (above) prediction using 193~{\AA} data, with the colormap corresponding to each row given on the right. The activation maps and the images have been rescaled between 0 and 1 row-wise for ease of comparison. For the fast wind prediction, note how the maximum activation occurs at the CH, 3 to 4 days prior to prediction, which seems to match with the correlations obtained in the literature ~\cite{CorHole_Vrsnak}. The slow wind, on the other hand activates the AR both closer and further away from prediction, and activated at the small CH on the closest day to prediction. However, other regions of the quiet Sun show a higher activation further away from the day of prediction. The slow wind activation is quite mixed and unclear when compared with the fast wind activation.}
\label{fig:GC}
\end{figure}
\begin{figure}[ht!]
\includegraphics[width=\linewidth]{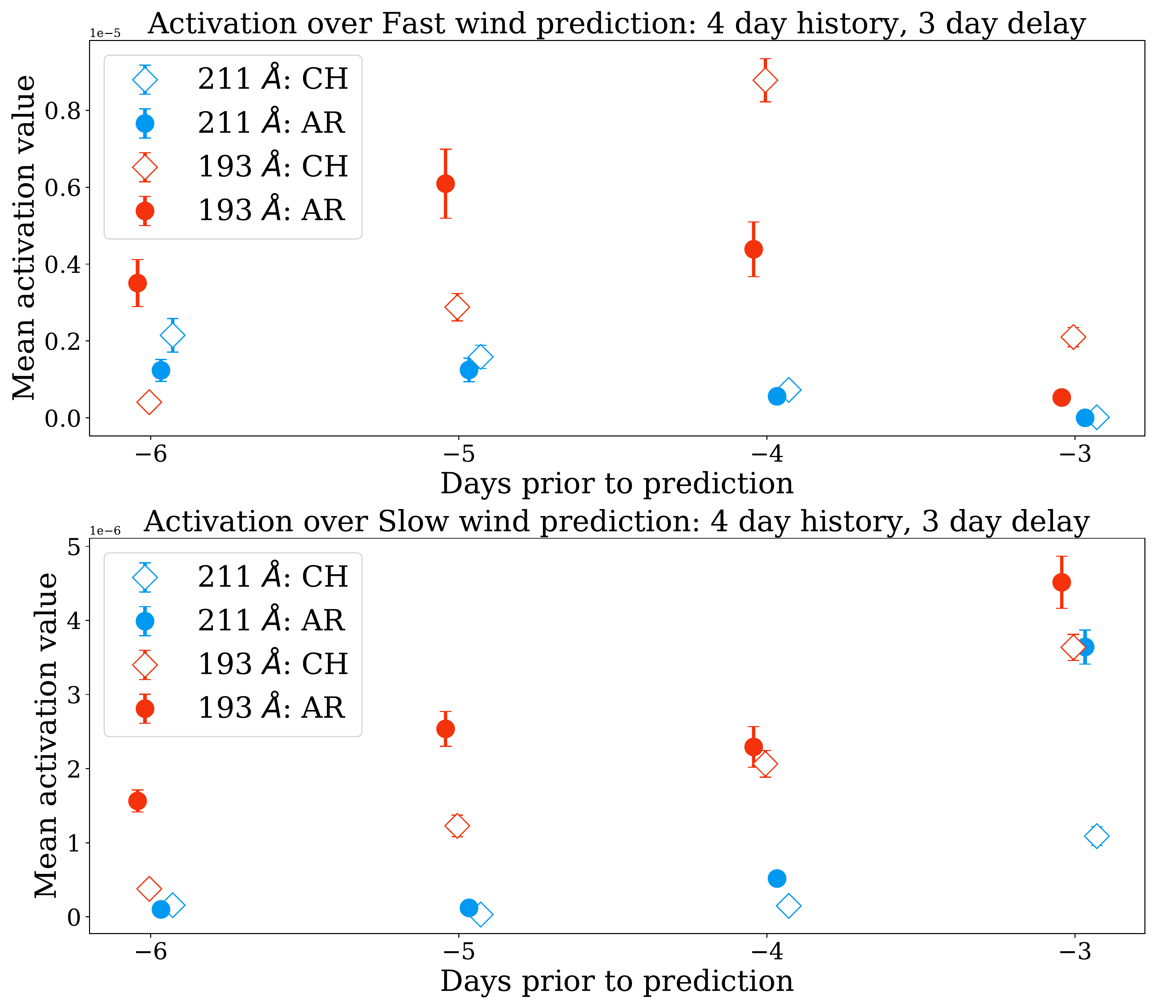}
\caption{Variation of mean activation for the 4 day history and 3 day delay model, for fast and slow SW prediction respectively. The activation is shown for models using 193{\AA} and 211{\AA} data respectively. The activation is shown over CH and AR alone. The errorbars indicate the standard error on the mean value, estimated from the standard deviation of the sample of activations. Please note that the errorbars here represent $3S$, i.e. thrice the standard error to make sure they are visible. Those activations with seemingly no errorbars have very small errors.}
\label{fig:att43}
\end{figure}
\begin{figure}[ht!]
\includegraphics[width=\linewidth]{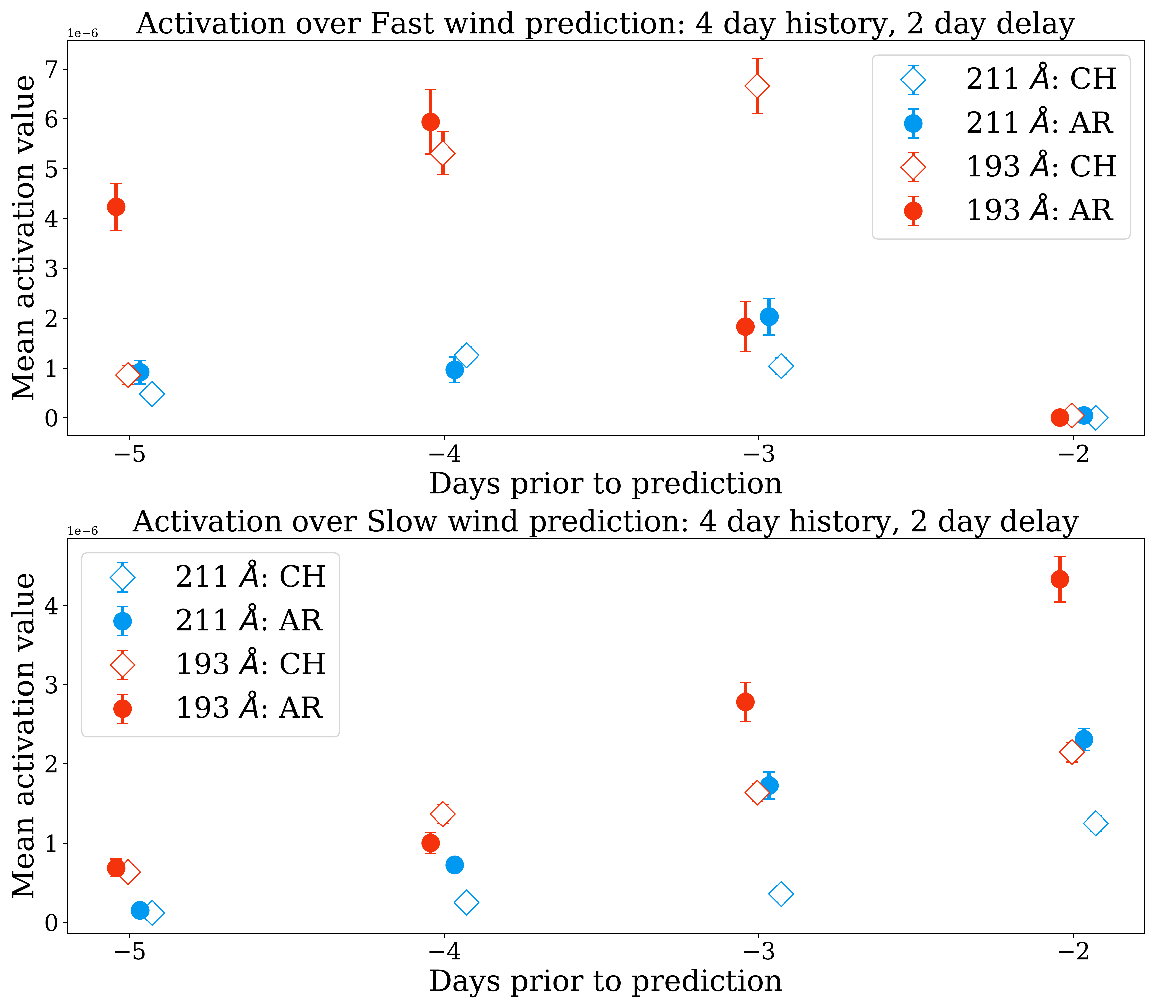}
\caption{Variation of mean activation for the 4 day history and 2 day delay model, for a fast and slow SW prediction respectively. The activation is shown for models using 193{\AA} and 211{\AA} data respectively. The activation is shown over CH and AR alone. The errorbars indicate the standard error on the mean value, estimated from the standard deviation of the sample of activations. Please note that the errorbars here represent $3S$, i.e. thrice the standard error to make sure they are visible. Those activations with seemingly no errorbars have very small errors.}
\label{fig:att42}
\end{figure}

Using Grad-CAM activation maps (described in Sec.~\ref{sec:QGC}) to visualize the activation, we analyze the activation of our models for various days of data. Fig.~\ref{fig:GC2} shows a sample Grad-CAM map from a fast and slow wind prediction using 211~{\AA} data, and a similar map is shown from 193~{\AA} prediction in Fig.~\ref{fig:GC}, for comparison. We see that the CHs are activated for the prediction of the fast wind, and the ARs are predominantly activated for the slow wind prediction. The CH peak activation for fast wind occurs ~3 to 4 days prior to prediction, which seems to corroborate with the correlations independently obtained~\cite{CorHole_Vrsnak}. %For a wind speed of $600$ km/s, considering the distance from the Sun to L1 to be 148.5 million km, an approximate travel time of $3$ days is obtained, which also seems to agree with the peak activation 3 to 4 days prior to prediction.
The slow wind activation is peaked at the AR close to the day of prediction (and also at the earliest day prior to prediction for 211~{\AA} data), with activation at other regions of the Sun further away from prediction. We hypothesize this might be due to bias of the LSTM to the most recent input to the network -- but this is still a hypothesis. 

To understand the statistics of activation given to CHs and ARs, we look at the mean activation value (as described previously), and plot it for `fast-wind' and `slow-wind' predictions from the model. While each cross validation model set will have its own activation plot, we present the plot for both 193{\AA} and 211{\AA} models using 4 day history of data with 3 days of delay. We also plot the activation for the models using 4 day history of data with 2 days of delay, since it shows consistent (and good) performance using both 193 {\AA} and 211 {\AA} data. These trends are shown in Fig.~\ref{fig:att43} and Fig.~\ref{fig:att42} respectively.

It can be seen from these plots that the fast SW induces greater activation at the CHs closer to the day of prediction, and the activation (at CH) decreases as we go farther into the past -- however, for the 211~{\AA} model, the activation shows slight increase. The fast wind also seems to activate the AR at much further times -- for both 193~{\AA} and 211~{\AA}. Note, however, that the peak CH activation is larger than the peak AR activation for 193~{\AA} -- for 211~{\AA} data, they are consistent within the errors in Fig.~\ref{fig:att43}. For the same parameters in Fig.~\ref{fig:att42}, CH peaks at 3 days prior to prediction for both the channels, and then goes down. Interestingly, however, the AR also seems to be activated to a similar level, but much further away from prediction time.

For the slow wind, activation for ARs remains high for much longer duration than the CHs -- however, the peak occurs closer to the day of prediction, rather than further away from prediction. This trend is seen in both Fig.~\ref{fig:att43} and Fig.~\ref{fig:att42}.

%One way of understanding the significance of these activation values is to look at the correlation of our target slow wind and predictions, and compare with the target fast wind and our predictions. For a H=4, D=3 model, we obtain $0.233 \pm 0.085$ for fast wind and $0.100 \pm 0.059$ for slow wind, for 193~{\AA} model. However, for the 211~{\AA} model, we obtain correlations of $0.117 \pm 0.269$ for fast wind and $0.110 \pm 0.215$ for slow wind respectively. These correlations indicate our 193~{\AA} performs better for fast wind and not as well for slow wind, while 211~{\AA} performs fine (but there is a large spread) for both fast and slow wind. From this, we can qualitatively say the fast wind activation of 193~{\AA} are more reliable than the slow wind activation using same data. The activation for both fast and slow wind of 211~{\AA} are not as reliable, and hence should be taken cautiously. 
%The true cause of such a trend is difficult to explain due to a `black-box' nature of Neural Networks. Overall, since the correlation of the whole series itself lies near 0.55 (i.e, the best model), these activations must be approached cautiously. Thus, the correlation values obtained, or the goodness of fit defines the limit of extraction of physics out of these DL models. 
%-------------------------------------->
\section{Discussion}
\label{sec:discuss}

The problem of SW prediction can be approached in two ways -- one, through purely theoretical modelling of the mechanism and fine tuning parameters to fit observations, and two, through purely empirical modelling and attempting to extract the physics. %Our model is an attempt at having an `SW generator in a lab', for accurate prediction of the wind speed given imagery data. As a starting step, 
We propose the WindNet to empirically model SW speed using AIA imagery data alone. We are able to predict the SW speed better with the 211~{\AA} data, and obtain a correlation of $0.55$ with the observed wind speed in the cross validation. The best performing models using 193~{\AA} and 211~{\AA} both outperform most of the larger delay benchmark models, and the 27-day persistence model. The $\chi^2_{red}$, which accounts for uncertainty in the measurement itself, indicates that our best models outperform the 27-day persistence,and are only slightly worse off than an autoregressive model with a single day delay -- more so for lead time predictions of 3-4 days. \\
The Activation plots seem to suggest the trained WindNet model pays attention to certain solar features consistent with heuristic expectations from solar wind theory, ex. %to indicate some physics of solar wind generation being captured by our model -- especially
the peak in activation at the CH 3-4 days prior to prediction. However, the significance of interpretation of activation values depend on a couple of other factors: 
\begin{itemize}
    \item Fitting error of our model: We still have a maximum correlation of 0.51(0.55) for the 193~{\AA}(211~{\AA}) data. 
    \item Visualization: The Grad-CAM used in this work gives a very coarse localization of activation, and thus may not point to precise origin of the particular kinds of wind. 
    \item Segmentation: Defining a region as CH accurately is difficult with intensity values alone -- ideally, one would require extrapolated magnetic field lines to check for CHs. Thus, an accurate definition of CHs based on intensity is required. In this work, we attempted to automate both the CH and AR definitions using histogram analysis automatically, thus there is bound to be some form of uncertainty. Hence, better segmentation methods may accurately capture the entire activation within a CH or an AR, and give a much better estimate of activation per unit area. 
\end{itemize}

At first glance, our WindNet may appear to not outperform existing models in terms of the metrics used. However, comparing our model to existing models (like the regressive models of \cite{MainRefRotterSW}, or \cite{wang1990solar}) would be an apples-to-oranges comparison, since:
\begin{itemize}
    \item We perform predictions over multiple Carrington rotations on the whole 8-year dataset.
    \item Our prediction target is the daily averaged SW speed, and as such must be compared to daily averaged predictions by other models. .
    \item We perform 5-fold cross validation on this dataset. However, due to a lack of confidence intervals in the previous results, we are unable to check if our results are statistically different from the existing models.
    \item Our model is built entirely using open source software, and the codebase may be used by the community at large to improve upon the results.
\end{itemize}

Thus, any benchmarking of our model must be done with models undergoing the same data preparation procedure, the same span of data and at the same cadence. We thus do not compare our results with the existing aforementioned models.

To overcome this limitation, we propose empirical benchmark models not unlike the existing empirical solar wind prediction models. In this regard, WindNet shows reasonable performance vis-a-vis the benchmark models -- however, there are numerous improvements possible.
 \begin{itemize}
    \item Data preparation: As H+D increases, more samples are discarded (as explained in Sec.~\ref{subsec:ControlParam}). This may be made more efficient by performing the Cross Validation (CV) first, and then splitting into folds later with the downside of high memory consumption.
    \item ICME mitigation: Our random assignment of 5-fold cross validation is to ensure the ICMEs are distributed uniformly across all the folds, thereby influencing all the CVs equally. Due to inadequate number of ICME samples, we do not characterise them.
    \item Network architecture: Better architectures may be designed to improve the prediction vis-a-vis the observations, or more novel ML methods may be employed for a direct prediction.
    \item Visualization: Visualization of ML models is an hot area of research in the ML community -- thus, more accurate visualization techniques may be expected to emerge in coming years.
    \item TS evaluation: As seen in Sec.~\ref{sec:result}, the HSE capturing algorithm misses many potential enhancements due to the speed increases not satisfying the absolute speed change criteria. Hence, the TS evaluation should be taken with caution. 
 \end{itemize}

This work is a first step toward training and testing various ML models for predicting other SW target parameters, such as proton density, temperature and  magnetic field (specifically, $B_z$). The code and data used in this work is open-sourced (model may be found on github: \url{https://github.com/Vishal-Upendran/WindNet}). Our publicly released source code promotes reproducible research by allowing others to reproduce the results presented here. This includes data partitioning, cross-validation, model training, and evaluation. This code base can be built upon by other researchers to further improve the performance of solar wind prediction models. Furthermore, with the ever-increasing research on Interpretable AI, this codebase may be used by researchers to come up with various methods of visualizations to quantify regions of solar wind emergence.

%------------------------------------->
\acknowledgments
We acknowledge use of NASA/GSFC's Space Physics Data Facility's OMNIWeb service, and OMNI data, and the AIA data is available on the Stanford Digital repository. U.V. would like to thank Alex Varghese and Mahendra Khened (Medical image Reconstruction Laboratory, Dept. of Engineering Design, IIT Madras) for long discussions on activation visualization and Interpretable ML. U.V would also like to thank Dattaraj Dhuri (Dept. of Astronomy and Astrophysics, TIFR-Mumbai) for independent verification of codes and results. U.V. would like to thank Durgesh Tripathi (Inter University Centre for Astronomy and Astrophysics, Pune, India) for providing computing facility, and insightful comments on the statistics of results. The authors would also like to acknowledge the two anonymous refrees, whose comments helped substantially improve the manuscript. SH and U.V acknowledge support from the Max-Planck Partner Group Program and the Ramanujan Fellowship SB/S2/RJN-73. M.C.M.C acknowledges support from NASA's SDO/AIA (NNG04EA00C) contract to LMSAL. AIA is an instrument onboard SDO, a mission for NASA's Living With a Star program.

%% ------------------------------------------------------------------------ %%
%% References and Citations

%%%%%%%%%%%%%%%%%%%%%%%%%%%%%%%%%%%%%%%%%%%%%%%
%
% \bibliography{<name of your .bib file>} don't specify the file extension
%
% don't specify bibliographystyle
%%%%%%%%%%%%%%%%%%%%%%%%%%%%%%%%%%%%%%%%%%%%%%%

\bibliography{Referencefiles}

%Reference citation instructions and examples:
%
% Please use ONLY \cite and \citeA for reference citations.
% \cite for parenthetical references
% ...as shown in recent studies (Simpson et al., 2019)
% \citeA for in-text citations
% ...Simpson et al. (2019) have shown...
%
%
%...as shown by \citeA{jskilby}.
%...as shown by \citeA{lewin76}, \citeA{carson86}, \citeA{bartoldy02}, and \citeA{rinaldi03}.
%...has been shown \cite{jskilbye}.
%...has been shown \cite{lewin76,carson86,bartoldy02,rinaldi03}.
%... \cite <i.e.>[]{lewin76,carson86,bartoldy02,rinaldi03}.
%...has been shown by \cite <e.g.,>[and others]{lewin76}.
%
% apacite uses < > for prenotes and [ ] for postnotes
% DO NOT use other cite commands (e.g., \citet, \citep, \citeyear, \nocite, \citealp, etc.).
%

\end{document}